\documentclass[iop]{emulateapj}
\usepackage{multirow}
\usepackage{natbib}
\usepackage[latin1]{inputenc}
\usepackage{amsmath}
\usepackage{rotating}
\usepackage[table]{xcolor}
\usepackage{color,soul}
\usepackage{hyperref}

\newcommand{\myemail}{martina.fagioli@phys.ethz.ch}

\shortauthors{Fagioli et al.}

\begin{document}

\title{Minor Mergers or Progenitor Bias? The Stellar Ages of Small and Large Quenched Early-Type Galaxies}

\author{Martina Fagioli\altaffilmark{1}, C. Marcella Carollo\altaffilmark{1}, Alvio Renzini\altaffilmark{2}, Simon J. Lilly\altaffilmark{1}, Masato Onodera\altaffilmark{1,$\dagger$} and Sandro Tacchella\altaffilmark{1}}

\altaffiltext{1}{Institute for Astronomy, ETH Zurich, 8093 Zurich, Switzerland;  \myemail}
\altaffiltext{2}{INAF-Osservatorio Astronomico di Padova, Vicolo dell'Osservatorio 5, I-35122, Padova, Italy}
\altaffiltext{$\dagger$}{Current address: Subaru Telescope, National Astronomical Observatory of Japan, National Institutes of Natural Sciences (NINS), 650 North A'ohoku Place, Hilo, HI 96720, USA, and also Graduate University for Advanced Studies, 2-21-1 Osawa, Mitaka, Tokyo, Japan}

\begin{abstract}
We investigate the origin of the evolution of the population-averaged size of quenched galaxies (QGs) through a spectroscopic analysis of their stellar ages. The two most favoured scenarios for this evolution  are either  the size growth of individual galaxies through a sequence of dry minor merger events, or the addition of larger, newly quenched galaxies to the pre-existing population (i.e., a progenitor bias effect). We use the 20k zCOSMOS-bright spectroscopic survey to select \textit{bona fide} quiescent galaxies at  $0.2<z<0.8$. We stack their spectra in bins of redshift, stellar mass and size to compute stellar population parameters in these bins through fits to the rest-frame optical spectra  and through Lick spectral indices. We confirm a change of behaviour in the size-age relation below and above the $\sim10^{11} \mathrm{M}_\odot$ stellar mass scale: In our $10.5 < \log \mathrm{M_*/M_\odot} < 11$ mass bin, over the entire redshift window, the stellar populations of the largest galaxies are systematically younger than those of the smaller counterparts, pointing at progenitor bias as the main driver of the observed average size evolution at sub-10$^{11} \mathrm{M}_\odot$ masses. In contrast, at higher masses, there is no clear trend in age as a function of galaxy size, supporting a substantial role of dry mergers in increasing the sizes of these most massive QGs with cosmic time. Within the errors, the [$\alpha$/Fe] abundance ratios of QGs are $(i)$ above-solar over the entire redshift range of our analysis, hinting  at universally short timescales for the buildup of the stellar populations of QGs, and $(ii)$ similar at all  masses and sizes, suggesting similar (short) timescales for the whole QG population and strengthening the role of mergers in the buildup of the most massive QGs in the Universe.
\end{abstract}

\section{Introduction}

The observed evolution with cosmic time in the population-averaged size of Quenched  Galaxies (QGs, here often also referred to as `passive' or `quiescent' galaxies, as opposed to `star-forming' galaxies) at fixed stellar mass has received a lot of attention in the past decade (e.g., \citealt{daddi2005, trujillo2007, cimatti2008, vandokkum2008, cassata2011, carollo2013}, hereafter C13, \citealt{poggianti2013, vanderwel2014}). The median half-light radius of QGs is about a factor $\sim$3--5 larger in the local universe than at redshift $z\sim 2$ \citep{newman2012}. The size growth scales as roughly $(1+z)^{-1}$, and it is similar to the rate of growth of the sizes of dark matter halos, but is somewhat steeper than the latter. This has sparked an intense debate concerning the physical mechanism behind this size evolution.

There are two main scenarios to which the evolution of the size-mass relation has been ascribed: the growth of individual QGs through a series of dry minor merger events, or the continuous addition of larger, recently quenched, galaxies, at later epochs.  This effect is an example of so-called `progenitor bias' in the sense that the population changes because of a change in membership rather than through changes in individual members (e.g., \citealt{franx2008, newman2012}; C13; \citealt{poggianti2013, belli2015}).

In the individual growth scenario, the compact cores of QGs would remain constant in mass within a few kiloparsecs, but would accrete extended stellar envelopes around them (\citealt{cimatti2008,hopkins2009, naab2009, cappellari2013a}). Contrary to major mergers, minor gas-poor (dubbed `dry') mergers could have a key role: for a given amount of added mass, mergers with higher mass-ratios (i.e. minor mergers) result in a larger size increase (\citealt{villumsen1983, hilz2012}; see also \citealt{taylor2010, feldmann2010, szomoru2011, mclure2013, vanderwel2014}).  This scheme would require $\sim 10$ dry mergers with $\sim 1:10$ mass ratio to account for the observed growth in size  \citep{naab2009,vandesande2013}. The mergers are required to be dry since `wet' mergers, involving  gas-rich galaxies, are expected to lead to central star formation and therefore to a reduction of the half light radius of the primary galaxy.  At a mass ratio of 1:10, the companion of a $10^{11}\mathrm{M}_\odot$ galaxy is a $10^{10}\mathrm{M}_\odot$ galaxy. Galaxies of this mass are generally gas-rich systems (e.g., \citealt{santini2014, genzel2015}). Therefore, the sequence and number of the required dry mergers, without a substantial contribution of wet mergers, is quite problematical, an aspect which has been substantially ignored so far.

Regardless of whether the merger scenario can explain the observed effect, the possible effects of progenitor bias must anyway be taken into consideration.    An implicit assumption of the individual size growth view is that galaxies which are being quenched at different epochs have  similar properties. If not, however, this could lead to a progenitor bias effect. In the context of the evolution of the average size-mass relation,  the addition of newly quenched galaxies to a pre-existing population of QGs  could lead to an observed growth of the average size of the population even if individual early-type galaxies do not grow at all. This is particularly important especially in the light of the observed increase by about one order of magnitude of the comoving number density of massive (i.e., $\gtrsim10^{11}\mathrm{M}_\odot$) QGs   from $z=2$ to the present epoch  (e.g. \citealt{ilbert2010, cassata2013, muzzin2013}). 

Tracing the evolution of the number density of QGs of different sizes offers clues towards discriminating between the two scenarios. Different studies well agree on the evolution of the number densities of  the smallest and densest QGs at stellar masses above $\sim10^{11} \mathrm{M}_\odot$, where a steady decrease is observed with cosmic time. At lower masses, however, different authors report different results. For example, C13 did not find any change in the number density of their `compact' galaxies at masses $10.5 < \log\mathrm{M}_*/\mathrm{M}_\odot < 11$; they report instead a substantial  increase in the number density of large QGs.  The constancy of the compact population and the increase in the large population led those authors to   advance  the progenitor bias interpretation. At similar $10.5 < \log\mathrm{M}_*/\mathrm{M}_\odot < 11$ masses, however, \cite{vanderwel2014} report a strong decrease in the number density for compact QGs since $z=1.5$, and therefore interpret  their observed  disappearance of these objects at the lower redshifts  as indication of a growth in  size of individual QGs.  

In comparing results from different studies, it is important however to note that the adopted definition of stellar mass is an important factor when discussing the evolution of the size-mass relation.  In C13, and also in this paper, we will define the stellar masses to be the integral of the star formation rate (SFR). These are about 0.2 dex larger than the commonly used definition which subtracts the mass returned to the interstellar medium, i.e. the mass of surviving stars plus compact stellar remnants.  The former has the feature of remaining constant after the galaxy ceases star-formation, whereas the latter continually decreases. Thus, when comparing the properties of quenched galaxies in a given mass bin across cosmic time, one should clearly use the former. This effect  explains  part of the discrepancies found in the different number density analyses: effectively high redshift galaxies are given a spuriously high mass, which leads to them appearing to be too small and to have a higher number density at high redshift. Another factor that leads to different estimates for the evolution of the number densities of small QGs is the definition of the bins in which the densities are computed, in particular whether a single size threshold is used to compare number densities at different redshifts, or whether the bins are defined along the size-mass relation at each given redshift (which, due to its evolution, implies a comparison between populations of different sizes).

Number densities alone however   are not conclusive. C13 and     \citet{damjanov2015a} agree that, at masses below the $\sim10^{11} \mathrm{M}_\odot$ scale,  the number densities of compact QGs remain constant since at least $z \sim 1$; these authors reached however different conclusions  on the origin of this constancy.  \citet{damjanov2015a} proposed that the compact QG population is continuously replenished with younger members, so as to compensate for the shift towards larger sizes of individual galaxies  due to mergers. In contrast, C13 argued that the compact population remains stable since $z\sim1$ and the newly-accreted memberd of the population have increasingly larger sizes at steadily lower redshifts. 

These two interpretations can be easily tested through the average ages of the populations involved.   If the increase of the median size is due to the addition of newly-quenched galaxies that are progressively larger  towards lower redshifts, then, at any epoch, the  stellar populations of larger QGs should be \emph{younger} than those of smaller QGs of similar mass.
 On the other hand, if individual QGs  grow their sizes through mergers  and the number density of compact QGs remains more or less constant due to the continuous production of compact QGs, then, at any epoch, \emph{smaller} QGs should be younger on average than their larger relatives of similar mass. Therefore, the stellar ages of the galaxies offer a powerful discriminant between these two scenarios (see e.g. also \citealt{onodera2012, belli2014b, keating2015, yano2016}).  

C13 did a study of the colors of compact and large $<10^{11} \mathrm{M}_\odot$  QGs at different redshifts, and found that, at any epoch,   larger QGs appear to be bluer than those of their smaller counterparts; it is this result that led those authors to conclude  that, at these masses, the stellar populations of larger QGs are younger  than those of smaller QGs and thus that  the evolution in size of the whole populations is to  a large extent ascribable to the addition of recently quenched, larger QGs.  Galaxies quenched at later epochs are indeed expected to have larger sizes than the ones quenched earlier as (progenitor) star-forming galaxies also experience an evolution in their average size with cosmic time (e.g., \citealt{newman2012}).  

Stellar ages determined on the basis of a single rest-frame optical color (as done in C13), however, heavily suffer from the well-known degeneracy between age, metallicity and also, possibly most  problematically, dust effects  \citep{worthey1994}. We therefore push here the analysis of the stellar ages of small and large QG below and above the evidently important mass threshold of $10^{11} \mathrm{M}_\odot$ using more robust spectroscopic measurements of stellar population properties.  Our primary goal is to test whether and to what extent progenitor bias is driving the increase of the average size of passive galaxies as a function of stellar mass; specifically, we use  two mass bins with boundaries $10.5 < \log\mathrm{M}_*/\mathrm{M}_\odot < 11$  and $11 < \log\mathrm{M}_*/\mathrm{M}_\odot < 11.5$. 
We also use the spectroscopic diagnostics to study the ratio of different elements in the  attempt to constrain the timescales of buildup of the stellar populations of quenched galaxies of different masses and sizes. 

Even spectroscopically, however, residual degeneracies between effects of age and metallicity continue afflicting galaxy ages, which are not straightforward to obtain. In the last few years, a number of `full spectral fitting' codes (e.g, \citealp{ocvirk2006b, ocvirk2006a, koleva2009, cappellari2004}, STECKMAP, ULySS and pPXF, respectively) have been developed in order to address this issue. In the full spectral technique, a set of templates is used to fit the overall shape of the spectrum. The most recent full spectral fitting codes do not fit the overall shape of the continuum, thereby avoiding common problems such as flux calibration and extinction. Instead, a polynomial function is used to fit the shape of the continuum.
Full spectrum fitting codes are good at handling the impact of the age-metallicity degeneracy as they maximize the information used from the whole observed spectrum (\citealt{koleva2008, sanchezblazquez2011, beasley2015, ruizlara2015}). We therefore adopt this methodology to derive our fiducial stellar population ages in this paper.
 
Besides the full spectral fitting analysis, we have also used however the Lick line-strength indices to get independent estimates of ages and metallicities. The Lick system of spectral line indices is a commonly used method to determine ages and metallicities of stellar populations (e.g., \citealt[][]{burstein1984, gonzalez1993, carollo1994, worthey1994, worthey1997, trager1998, trager2000, trager2005, korn2005, poggianti2001, thomas2003a, thomas2003b, korn2005, schiavon2007, thomas2011, onodera2012, onodera2014}). 
The system consists of a set of 25 optical absorption line indices, spanning a wavelength range from $\sim$ 4080 to $\sim$ 6400 \AA{}. The absorption features are particularly useful because they are largely insensitive to dust attenuation \citep{macarthur2005}.
Even so, age-dating based on the Lick indices is not free from degeneracy effects. Its main pitfall is that most indices are sensitive to all the basic population parameters, namely age, metallicity and the ratio of $\alpha$-elements  to iron.  Circulation of the errors can generate spurious correlations or anticorrelations \citep{kuntschner2001,thomas2005,renzini2006}. For example, an underestimate in the strength of a Balmer line (mainly sensitive to age), due e.g. to partial filling in by an emission line, may lead to an overestimate in the age, but having an overestimated age, the procedure is forced to underestimate  metallicity in order to match the strength of the metal lines. In this way, a spurious age-metallicity anti-correlation can be generated. A strength of our analysis is thus to attempt to mitigate the intrinsic degeneracies by using and comparing for cross-validation both methodologies, i.e., the full spectral fitting approach and the Lick indices approach.

We include in our study a brief introspection of the $\alpha$-elements to $Fe$-elements abundance ratios  (i.e., [$\alpha$/Fe]) in QGs of different masses and sizes. The   [$\alpha$/Fe] ratio is a well-known diagnostics to constrain  formation timescales (\citealt{matteucci1986, pagel1995}), since $\alpha$-elements such as O, Ne, Mg, Si, S, Ar, Ca, and Ti (i.e., nuclei that are built up with $\alpha$-particles) are delivered mainly by core collapse (CC) supernova explosions of massive  stars and thus on much shorter timescales than elements such as Fe and Cr, which come predominantly from the delayed explosion of Type Ia supernovae (e.g., \citealt{nomoto1984, woosley1995, thielemann1996}). Enhanced values of [$\alpha$/Fe] ratio indicate a short formation timescale. \\

The paper is organized as follows. In Section \ref{dataset} we describe the data set, the zCOSMOS-bright 20k catalog, and its features. Section \ref{sampselmeasurements} summarizes the basic measurements and presents the spectroscopic sample selection in detail. Section \ref{analysis} presents the steps we took in the course of our analysis. Section \ref{sizemass} describes the binning in mass, size and redshift, and Section \ref{stacking} describes the stacking procedure that was used to obtain average spectra as a function of redshift, mass and size.  The fitting used to correct for the emission lines contribution is described in Section \ref{emcorrection}. Section \ref{fullspectralfitting} describes how we derived ages with full spectral fitting using \texttt{pPXF} and Section \ref{lickmeasurements} describes how we measured the Lick strengths and derived the stellar population parameters from them. In Section \ref{results} we present our results, followed by a discussion in Section \ref{discussion}. In Section \ref{conclusion} we summarize our paper and present the conclusions.\\

Through this paper we adopt a $\Lambda$-dominated Cold Dark Matter ($\Lambda$CDM) cosmology, with $\Omega_m\,=\,0.3$, $\Omega_\Lambda\,=\,0.7$ and $H_0\,=\,70$ km\,s$^{-1}\,$Mpc$^{-1}$. All magnitudes are given in AB system. We use `dex' to refer to the anti-logarithm, so that 0.3 dex represents a factor of 2.

\section{Data Set}\label{dataset}

We use the zCOSMOS-bright 20k sample, to which however we apply several cuts in order to limit the redshift range to the $0.2 \le z \le 0.8$ interval, the mass range to the $10.5 < \log\mathrm{M}_*/\mathrm{M}_\odot < 11.5$ interval, and to ensure that the selected galaxies are \textit{bona fide} passive systems. In Figure~\ref{fig: sampsel} we show a  schematic view of the selection criteria that we have applied to the original sample, and in the next sections we describe in detail the steps which lead to the final selection.

\subsection{The 20k zCOSMOS-bright catalog}

The spectra we employed come from the full zCOSMOS-bright 20k catalog \citep{lilly2007, lilly2009}. Here we briefly summarize the data. zCOSMOS-bright consists of about 20,000 galaxy spectra selected to have $I_{\rm AB}<$ 22.5 across the full 1.7 deg$^2$ in the COSMOS field \citep{scoville2007}. The zCOSMOS project \citep{lilly2007} is a large redshift survey of galaxies undertaken on the ESO VLT.  The bright part uses the VIMOS MR grating with a resolution of $R\sim600$ and a pixel size of $\approx2.553$ \AA{}. 

The VIMOS wavelength coverage spans from 5500 to 9700 \AA{}. The greatest advantage of the zCOSMOS spectroscopic survey is to combine high quality spectra with a compilation of multi-wavelength imaging of the COSMOS survey data set, including \textit{HST}/ACS data \citep{Koekemoer2007}. Therefore we have available high quality spectroscopic redshifts, photometrically derived quantities and high resolution images. 

The typical redshift uncertainty in zCOSMOS-bright is $\pm$ 110 km s$^{-1}$ \citep{lilly2007}. A confidence class parameter is introduced to estimate the reliability of the redshift assignment. Also, objects flagged with confidence class 4 usually show high quality spectra. The spectroscopic redshifts are compared with photometric redshifts derived from the COSMOS multi-band photometric data and a decimal number is used to flag the agreement or otherwise between the photometric and spectroscopic redshifts. For a complete description of confidence classes the reader is referred to \cite{lilly2007, lilly2009}.

\subsection{The S-COSMOS MIPS 24 $\mu$m catalog} 

The S-COSMOS survey \citep{sanders2007} is a deep infrared imaging survey, which comprises IRAC 3.6, 4.5, 5.8 and 8.0 $\mu$m and MIPS 24, 70 and 160 $\mu$m observations including the entire 2 deg$^2$ of the COSMOS-ACS field. It has been carried out with the Spitzer Space Telescope as part of the Spitzer Cycle 2 and 3 Legacy Programs. In Cycle 2, the COSMOS field has been mapped at 24 $\mu$m. The observations performed in Cycle 3 mapped  the entire COSMOS area reaching deeper flux limits, down to a flux density limit S$_{24\,\mu m}\approx$ 0.08 mJy. 

\section{Measurements and Sample Selection}\label{sampselmeasurements}

\subsection{Redshifts}

We selected objects within a restricted redshift range of $0.2<z<0.8$ so as to be complete in mass down to $\log\, \mathrm{M_*/M}_\odot =10.5$ at $z = 0.8$ \citep{pozzetti2010}. In order to achieve a high signal-to-noise ratio, adeguate for a Lick-based stellar population analysis, we further restricted the sample to confidence Classes 3 and 4 (including secondary objects in Classes 23 and 24, but excluding objects with broad emission lines, Classes 13 and 14). Class 3 and 4 spectra have a very secure redshift assignment (reliability $>99.8\%$).  Almost all the galaxies in our sample (98.5\%) have been flagged with the .5 decimal number, indicating an agreement between spectroscopic and photometric redshift to within $0.08(1+z)$ (a subsequent visual inspection of all of the final set of objects confirmed the correctness of the assigned spectroscopic redshifts).  
After this first selection, we end up with a sample of 9,208 objects.

\subsection{Stellar Masses}\label{stellarmasses}

Our sample is matched with that  from C13 to get structural parameters, whose derivation has been fully described in that paper. The stellar masses are derived with the \texttt{ZEBRA+} \citep{oesch2010} code from synthetic SED fitting to 11 photometric broad bands from 3832 \AA{} (u*, CFHT) to 4.5 $\mu$m (\textit{Spitzer}/IRAC channel 2). A set of star formation history models with exponentially declining SFRs, ranging metallicities from 0.05 to 2 Z$_\odot$, with decay timescales from $\tau\sim$ 0.05 to 9 Gyr, and ages from 0.01 to 12 Gyr, constitutes the SED library. This was constructed with the \cite{bruzual2003} (BC03) stellar population synthesis code with a Chabrier initial mass function \citep{chabrier2003}. We use dust reddening from \cite{calzetti2000} with $E(B-V)$ as a free parameter. As discussed in the Introduction, we stress that our stellar masses are defined as the time-integral of the SFRs, which are on average around 0.2 dex higher than stellar masses which exclude the mass that is subsequently returned to the interstellar medium.

\subsection{Sizes}

We adopt the half-light radius ($\mathrm{r}_{1/2}$) as an estimation of the size of our galaxies. The procedure is fully described in C13. The sizes of our galaxies have been measured with the software \texttt{ZEST+} (\textit{Zurich Estimator of Structural Types Plus}), an extended version of \texttt{ZEST} \citep{scarlata2007}. The main advantage is that it measures the half-light radii within elliptical apertures, instead of the circular apertures of \texttt{SExtractor} \citep{bertin1996}. \texttt{ZEST+} requires as input the total apparent flux of each galaxy, which is taken from the 2.5 R$_{\mathrm{Kron}}$ \citep{kron1980} value from \texttt{SExtractor}. Using a file with the positions of close objects, \texttt{ZEST+} first replaces the segmentation maps of companion galaxies with random sky values, and then estimates the local sky background of the galaxy. Finally, the code outputs the semi-major axis of the corresponding elliptical aperture that encloses half of the output total flux.  In Figure~\ref{fig: measurements} we show the stellar masses and sizes distribution before and after the mass cut.

\subsection{Spectroscopic Sample Selection}

\begin{figure*}
\centering
\includegraphics[width=160mm]{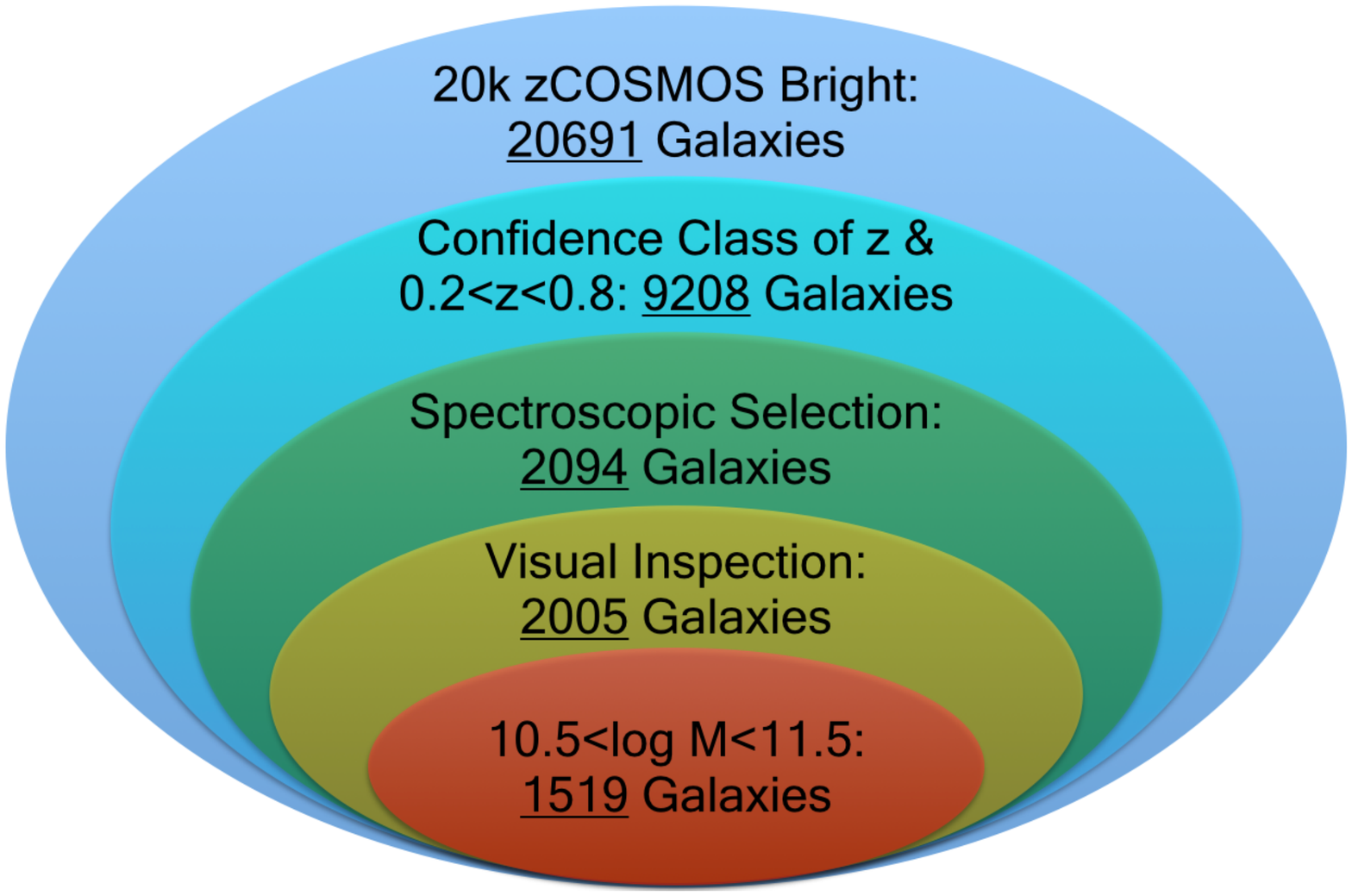}
\caption{Illustration of the steps taken in order to select the final sample. Starting from the full 20k zCOSMOS-bright catalog, we restrict the analysis to galaxies in the redshift range $0.2< z<0.8$ and to those with redshift confidence classes 3.1, 3.5, 4.1, 4.5, 23.1, 23.5, 24.1 and 24.5.  We then select purely passive galaxies by retaining objects with absent or very weak emission lines and excluding any objects with MIPS detections. In the case of a possible MIPS detection, a visual inspection was performed to see if the object in question was the infrared source. The analysis is restricted to 2 mass bins, $10^{10.5}<\mathrm{M/M_\odot}<10^{11}$ and $10^{11}<\mathrm{M/M_\odot}<10^{11.5}$, both of which are complete throughout the redshift range of interest.}
\label{fig: sampsel}
\end{figure*}

\begin{figure}
\centering
\includegraphics[width=88mm]{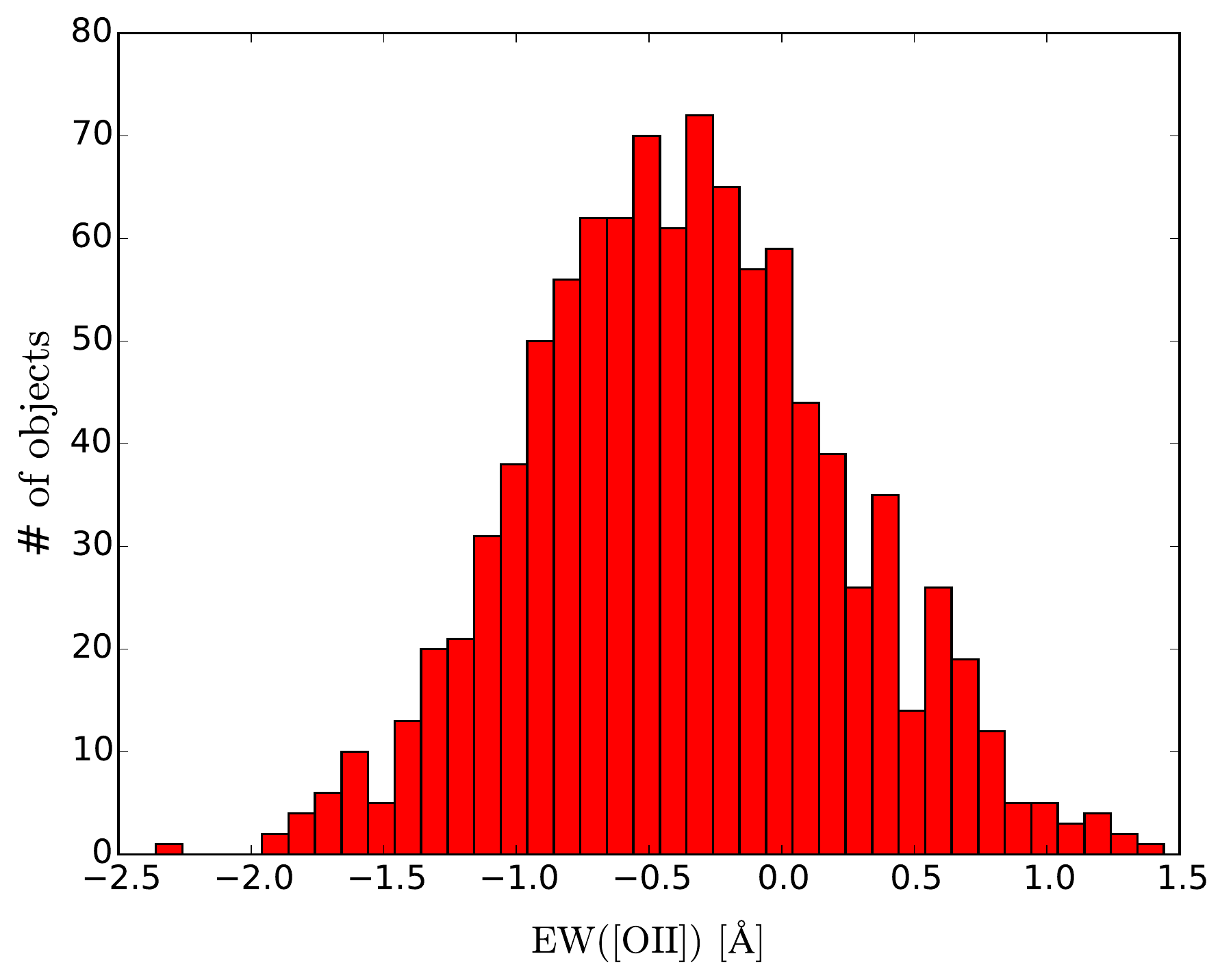}
\caption{Equivalent Widths of [OII] line (EW[OII]) measured on 1000 stacked spectra, each constructed from a random subsample of 100 of our visually inspected sample of galaxies. The definition of continuum and line bandpasses are from \cite{balogh1999}.  This tests the success in defining a set of galaxies that are free of emission-lines.  The typical stack shows a negative equivalent width, and the maximum ever seen is +1.5 \AA{}, well-below the 5 \AA{} limit which is commonly used in the literature to separate star-forming and passive galaxies (e.g., \citealp{mignoli2009, moresco2010}).}
\label{fig: equivwidth}
\end{figure}

\begin{figure*}
\centering
\includegraphics[height = 76mm]{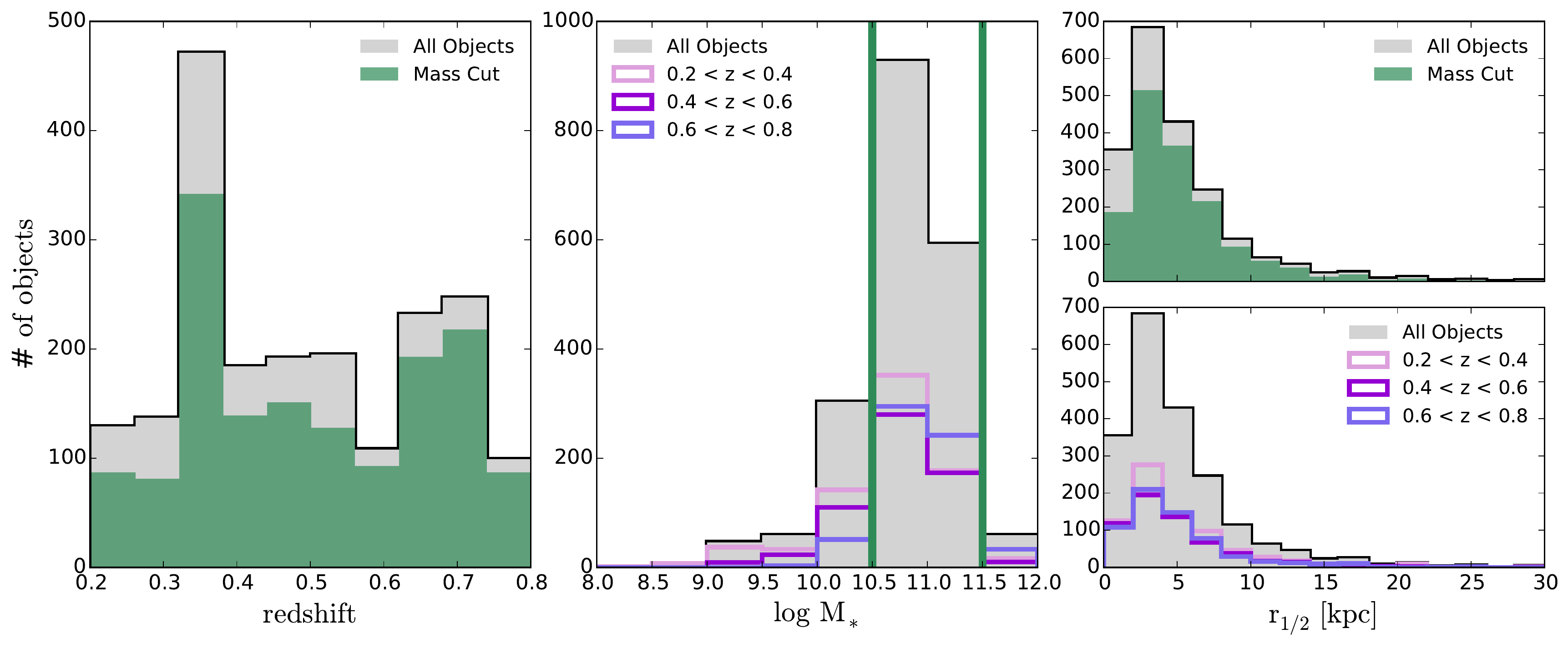}
\caption{Histograms showing the distribution of the sample with respect to redshift, mass and size, respectively. The left panel shows the redshift distribution of our objects. The central panel shows the distribution of stellar masses (i.e., integrated SFRs), measured with the \texttt{ZEBRA+} \citep{oesch2010} code. The different colors represent the 3 redshift bins our sample is divided in. The vertical dashed lines in the central panel show our mass cuts at $10^{10.5}\mathrm{M}_\odot$ and $10^{11.5}\mathrm{M}_\odot$. The right panel shows the size distribution in kpc. The size measurements and their correction have been described in C13.}
\label{fig: measurements}
\end{figure*}

To proceed towards our goal of studying the properties of QGs, we give special attention to the separation between star-forming and quiescent galaxies.  A number of studies have used  photometric information to achieve this separation; a UVJ color-color diagram has been found to be particularly effective in this direction (\citealt{wuyts2007, patel2009, williams2009, whitaker2011, muzzin2013, moresco2013}).  A cleaner separation can clearly be achieved however using spectroscopic diagnostics  (see e.g., \citealt{onodera2014}).   Our selection of quiescent galaxies was done by choosing objects which show no, or only very weak, emission lines, with the following procedure:
\begin{itemize}
\item [$\diamond$] Identify the expected wavelength of H$\alpha$, H$\beta$ and [OII] 3727 \AA{}.
\item [$\diamond$] Define a continuum region on either side of this location to be between 1 and 5 times the FWHM $\Delta\lambda$ of the instrumental resolution ($\mathrm{R}=\lambda/\Delta\lambda=600$) and compute the mean ($\langle f_{\mathrm{cont}}\rangle$) and the standard deviation ($\sigma$) of the continuum per pixel.
\end{itemize}
We then consider a line to be detected in emission if the peak of the line exceeds three times the noise per pixel,
\begin{equation}
\max\,(f_{\mathrm{line}}-\langle f_{\mathrm{cont}}\rangle)\, >\,3\sigma
\end{equation}
This is straightforwardly applied for  H$\alpha$ and [OII] and the object is excluded if either line is detected.  The case of $\mathrm{H}\beta$ is complicated by stellar absorption lines.  If neither of [OII] and H$\alpha$ lines is in the observed wavelength range, we use an empirical calibration of the ratio of the peak of the [OII] and $\mathrm{H}\beta$, [OII]$\simeq 2.8\mathrm{H}\beta-2$, which yields an equivalent threshold of $\gtrsim 1.8\sigma$, using the standard deviation of the continuum from the blue side of the feature only, due to the proximity of [OIII] 5007 \AA{}.  Our wavelength range included $\mathrm{H}\alpha$ for 1042 galaxies and [OII] 3727 \AA{} for 3826 objects. For the rest of the galaxies in the sample, we use H$\beta$ as the criterion to exclude emission line objects.
As a result of this selection, we find a total of 2,094 galaxies showing no detected emission lines.\\
\\
To further check against the presence of star forming objects, we cross-match these 2,094 galaxies with the S-COSMOS MIPS 24 $\mu$m Photometry Catalog (October 2008) and with C-COSMOS (Chandra COSMOS) \citep{elvis2009}, identifying sources within $2\arcsec$ (\citealp[see also][]{caputi2008}). We find 257 galaxies having a possible detection in the MIPS catalog and 28 in the X-ray. We discarded 38 objects which have a MIPS detection and for which subsequent visual inspection of the spectra revealed emission lines that had not been found by the automatic algorithm described above. The emission lines were hidden at the edges of the wavelength range, where the signal-to-noise ratio is lower and fringing may alter the shape of the continuum.  None of the objects with X-ray detection show emission features in the zCOSMOS spectra and therefore we keep those galaxies in the quiescent sample. 

As the selection of a purely quiescent sample is crucial for our analysis, because the star-forming objects are expected to be larger than quiescent ones, we visually inspected the spectra of all the galaxies in the sample. We further discard 51 galaxies for which visible emission lines had not been detected from the automatic code.  At the end of this selection, 2,005 objects have been defined as purely quiescent galaxies.  We stress that this selection has been made purely on the basis of the spectra (and MIPS and X-ray catalogues) with no reference to the images, sizes, or morphologies of the galaxies.

As a final check that we have succeeded in excluding all galaxies with significant emission lines, we stack the spectra of a subsample consisting of a randomly chosen $5\%$ of the final set of 2,005 galaxies and measure the Equivalent Width of [OII] $(\mathrm{EW([OII])}$), with continuum and line passbands defined as in \citealt{balogh1999}. We repeat this procedure 1000 times. Figure~\ref{fig: equivwidth} shows the distribution of the EW([OII]) of these stacked spectra. The EW([OII]) of the stacked spectra in every case far below the 5 \AA{} limit which is commonly used to separate star forming from quiescent galaxies (e.g., \citealp{mignoli2009, moresco2010}, see also \citealp{balogh1999} for the typical values of EW([OII]) they found for quiescent galaxies). In fact, the maximum EW for the [OII] emission line in these 1000 trials was $<2$ \AA{}, which corresponds roughly to a $\log\mathrm{(sSFR/Gyr}^{-1}) < -2$ \citep[see also Figure 5 in][]{moresco2013}, and the mean is below zero.

In Figure~\ref{fig: sampsel}, where we show the steps of our sample selection, we also give the number of galaxies remaining in the sample at each of the steps.  The final step is to apply a cut in stellar mass at $10.5<$ log M/M$_\odot <11.5$, which should be complete in the redshift range $0.2<z<0.8$, applying the procedure described in \citealt{pozzetti2010}.   Figure~\ref{fig: measurements} shows the distribution of the stellar masses of the passive sample, before the cut, and of the redshifts and sizes before and after the mass cut is applied.

\section{Analysis}\label{analysis}

\subsection{Sample Binning}\label{sizemass}

In order to study the average properties of galaxies in our sample, we split them into bins of redshift, stellar mass and size. 
Our galaxies have three equal bins in redshift within the redshift interval $0.2 < z < 0.8$, with $\Delta z=0.2$. We cut the redshift interval at $z =0.8$ in order to be complete in mass down to $\log\mathrm{M_*/M_\odot} =10.5$.  We then divide the mass range into two bins with size of 0.5 dex, as in C13,  and further divide the sample into three size bins, following two different procedures, as follows. 

\begin{figure*}
\centering
\includegraphics[height = 130mm]{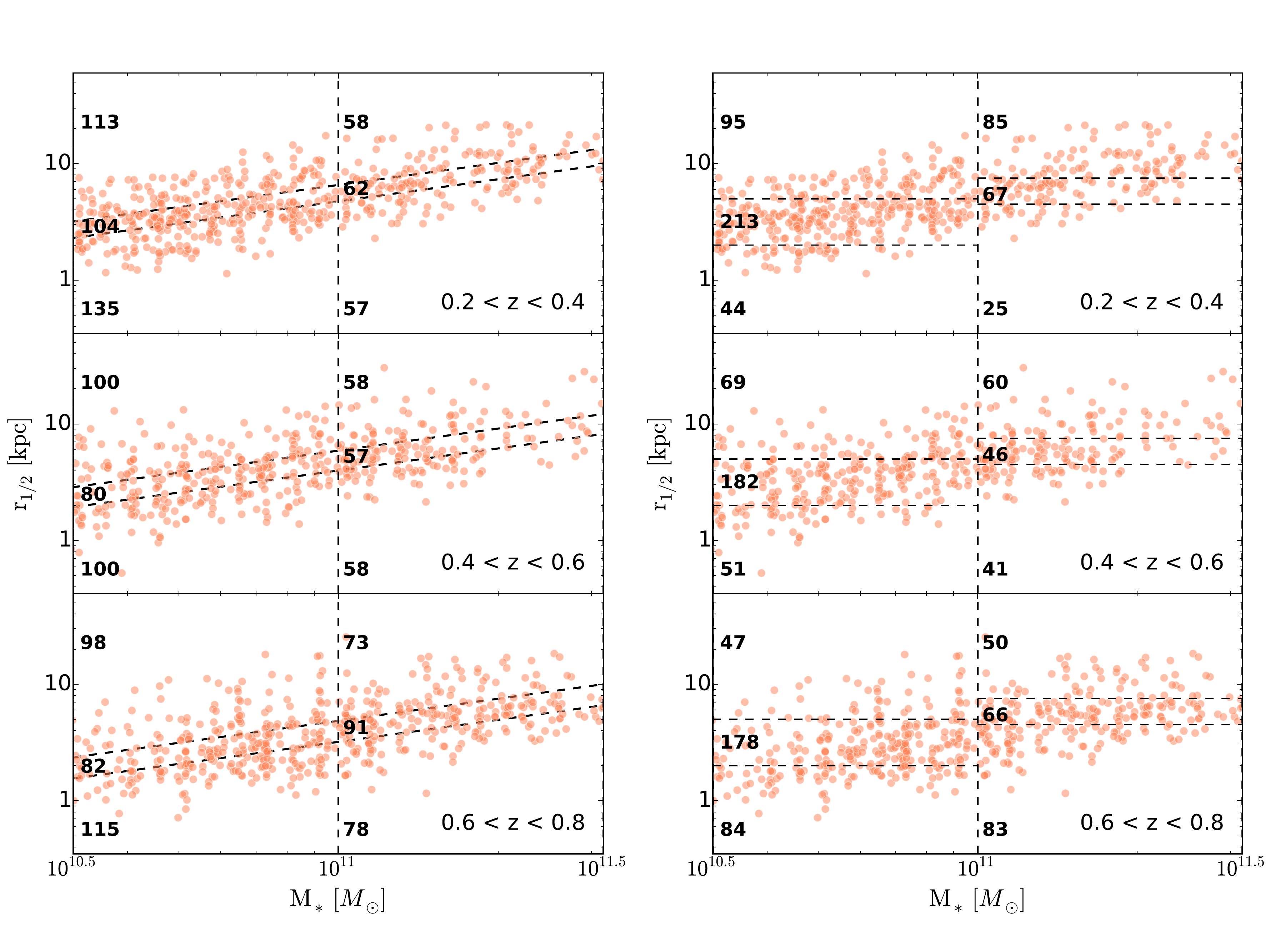}
\caption{The size-mass relation for our sample (red circles) of QGs. Dashed lines indicate the adopted bins in size and stellar mass in each redshift interval. The three panels in each plot show the three different redshift bins. Numbers in the boxes are the number of objects to be stacked in each redshift, mass and size bin. On the left, the two black dashed lines which follow the slope of the size-mass relation in each panel enclose the middle 30$\%$ of the distribution at each redshift.   These bins allow the comparison of galaxies of the same size relative to the evolving mean relation.  In contrast, the horizontal bins in the right hand panel are chosen to allow the comparison of galaxies with the same absolute size at different redshifts. }
\label{fig: number}
\end{figure*}

Quiescent galaxies are known to follow a tight relation between their size (r$_{1/2}$) and stellar mass (M$_*$), which evolves with redshift (\citealt{daddi2005, williams2010, newman2012, patel2013, mosleh2013, vanderwel2014, tacchella2015a}). Our final goal is to compare the average stellar population parameters for a sample of `small' and `large' QGs. Therefore, we fit a size-mass relation for the different redshift bins, finding minimal or no variation in the slope, as expected (e.g., \citealt{vanderwel2014}).  We therefore fix the slope at the value obtained in the central redshift bin, $0.4 < z< 0.6$, as shown in the central panel in the lefthand figure in Figure~\ref{fig: number}.  The resulting values of the intercepts at different redshifts are given in Table~\ref{fig: intercept}. As expected, we find decrease in the size (at given mass) with redshift.  Our first approach to binning in size constructs bins relative to this evolving mean relation. This is shown in the lefthand plot in Figure~\ref{fig: number}. The two dashed lines, with the same slope as the our fitted size-mass relation, split the sample into $35:30:35 \%$ of the galaxies (across the whole redshift range). We define the galaxies lying in these areas as `small', `intermediate size' and `large' galaxies.  The bold numbers in Figure~\ref{fig: number} show how many passive galaxies we stacked as a function of redshift, size and mass. We define this binning as `size-mass cut'.

\begin{table}
\caption{Intercept and slope in the size--mass relation as a function of redshift ($\log\,\mathrm{r}_{1/2} = \alpha\log\,\mathrm{M_*}-\beta$)}

\centering
\begin{tabular}{cc} 

   &  \\
Redshift  & Slope ($\alpha$) \\
  \hline
  \hline
    &  \\
 $0.2 < z < 0.8$ & 0.63 \\
 \hline
      &  \\

Redshift  & Intercept ($\beta$)\\
\hline
\hline
 &  \\
 $0.2 < z < 0.4$ & 5.76\\
  &  \\
 $0.4 < z < 0.6$ & 6.06\\
  &  \\
 $0.6 < z < 0.8$ & 6.43\\

\hline
\end{tabular}

\label{fig: intercept}

\end{table}

This cut is useful to compare the average stellar ages among different sizes in the same redshift bin. However, the comparison of stellar ages between different redshift intervals becomes difficult, as different populations of galaxies may enter into the definition of small and large as the mean relation evolves. To make such comparison, we also apply a different binning in size, which we define as `horizontal cut' (as shown in the righthand panels in Figure~\ref{fig: number}). In this case, we define for each mass bin, three bins in size that are the same at all three redshifts.  Specifically, for the mass bin $10.5<\log \mathrm{M}_*/\mathrm{M}_\odot<11$, we define as small the galaxies having $r_{1/2} < 2$ kpc, intermediate, $2<r_{1/2}<5$ and large, $r_{1/2}>5$ kpc. For more massive galaxies ($11<\log \mathrm{M}_*/\mathrm{M}_\odot<11.5$), small galaxies have r$_{1/2}<4.5$ kpc and large $>7.5$ kpc. This allows us to track the evolution of the stellar ages of galaxies with same size and same stellar masses (i.e., integrated SFRs, see Section~\ref{stellarmasses}) through different cosmic epochs.

\subsection{Stacking}\label{stacking}

\begin{figure*}
\centering
\includegraphics[width=158mm]{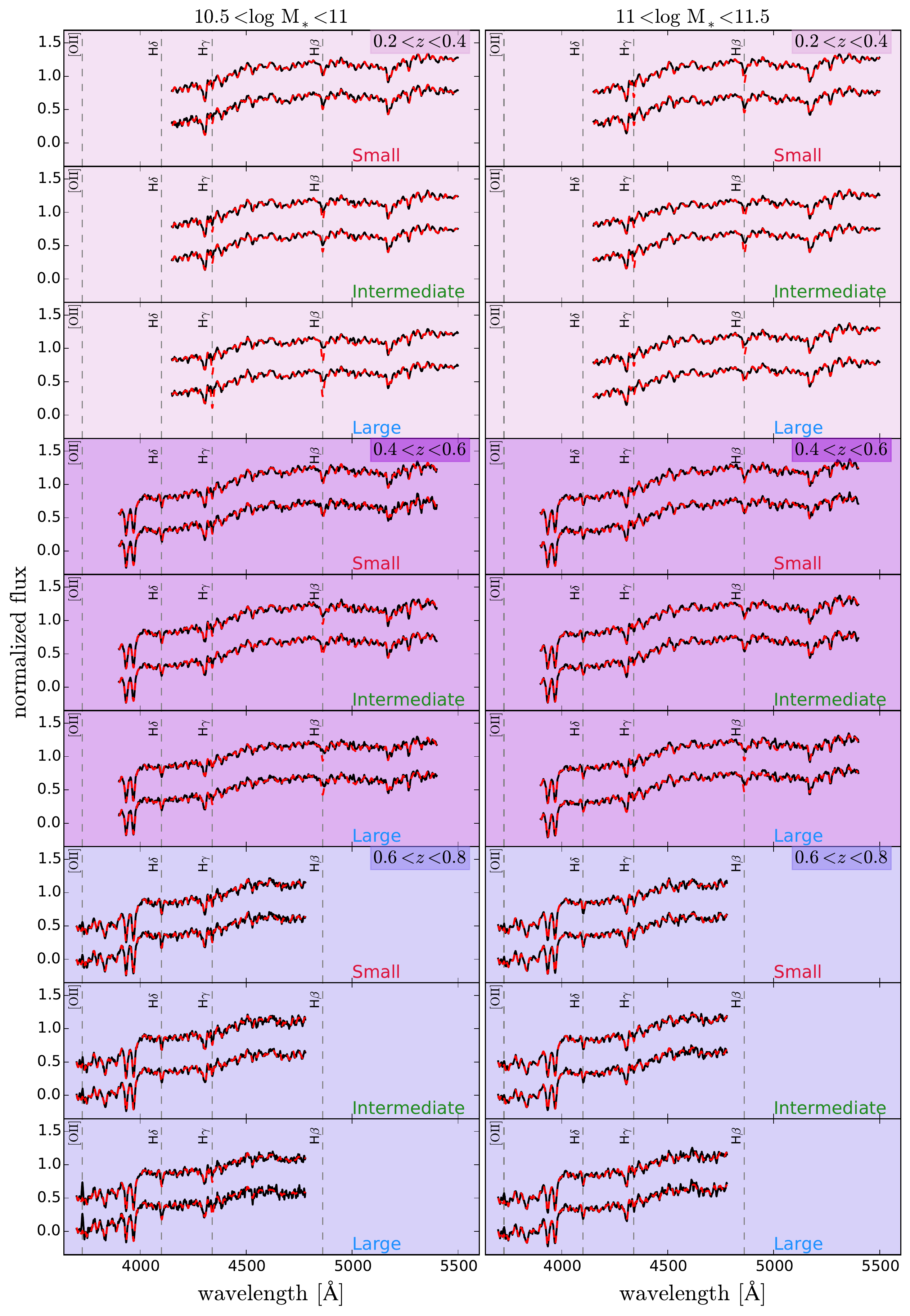}
\caption{Stacked spectra for all our redshift, mass and size bins. From top to bottom, the different shades of purple as background color separate the spectra in redshift. On the left side, we show galaxies with masses between $10.5 < \log\,\mathrm{M_*/M}_\odot< 11$, and on the right side, the spectra of more massive galaxies  ($11 < \log\,\mathrm{M_*/M}_\odot< 11.5$). The three size bins are as described in Section~\ref{sizemass} and shown in Figure~\ref{fig: number}.  In each panel, the upper spectrum shows the spectrum that is obtained by stacking spectra binned along the size-mass relation while the lower spectrum shows the stack obtained with the horizontal cut, shifted by subtracting a constant.  In each case, the black line shows the observed stacked spectra, while the red dashed line shows the best-fit template that is obtained with \texttt{pPXF}.}
\label{fig: allspectra}
\end{figure*}

\begin{figure*}
\centering
\includegraphics[height = 180mm]{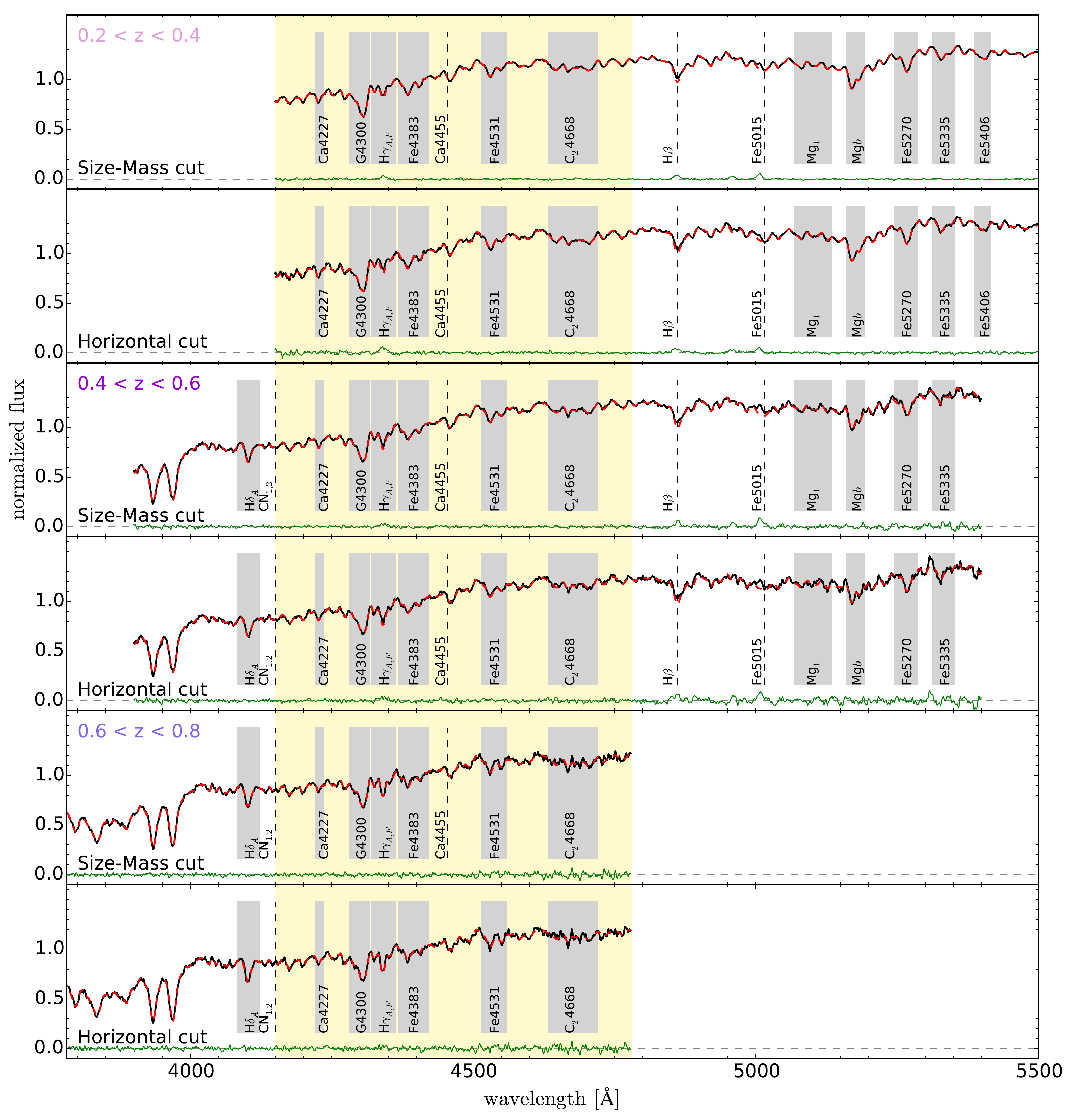}
\caption{Stacked spectra of the small galaxies with mass $10.5 < \log\,\mathrm{M}_*/\mathrm{M}_\odot < 11$ in three different redshift bins.   The shaded areas show the position of the absorption features that can be observed in each redshift bin.  The dashed black lines indicate the indices which have been excluded in the analysis following \cite{thomas2011}. The shaded yellow area shows the overlapping wavelength range between different redshifts. The black solid line is our stacked observed spectra and the dashed red line is the best-fit obtained with \texttt{pPXF} \citep{cappellari2004}. The residuals (shown as the solid green line along the bottom ) show the tiny contamination from emission line fill-in at the position of $\mathrm{H}\beta$ and [OIII] doublet. We discuss in detail the implications of these residuals in Section~\ref{emcorrection}}
\label{fig: stacked}
\end{figure*}

We computed the average stacked galaxy and noise spectra for the galaxies in each bin of redshift, stellar mass and size.  First, we de-redden the individual spectra for Galactic extinction following the extinction curve for  diffuse gas from \cite{odonnell1994} with $R_{\rm V} = 3.1$, and using the Galactic $E(B - V)$ values from the maps of \cite{schlegel1998}.  We find a typical correction factor of $f_{\mathrm{obs}}\sim0.96f_{\mathrm{dered}}$. We do not correct for any internal dust extinction in the galaxies, since they are expected to be passive systems with negligible dust.

We then de-redshift the spectra to the rest frame and normalize them by the mean flux at rest frame $4,100< \lambda < 4,700$ \AA. The spectra are then interpolated onto a 1 \AA{} linearly spaced wavelength grid. The associated noise spectra, which we use as weights during the stacking procedure, are normalized by the same factor as that for the object spectra and interpolated to the rest-frame in quadrature. The spectra are a straight average of the individual spectra, weighted by the signal to noise in each spectrum at each wavelength.

In Figure~\ref{fig: allspectra} we show the stacked spectra of all bins of our analysis, i.e., all redshifts, mass bins, and size bins. In each panel we show two spectra, respectively for the size-mass binning and for the horizontal-cut binning.  To highlight the quality of the stacked spectra, we replot in Figure~\ref{fig: stacked} the spectra of only the small galaxies, this time however with an expanded horizontal axis. The main absorption lines that we have used for our measurements of ages are marked with vertical grey bands; also shown, with vertical dashed lines, the spectral features which were excluded from the computation of the stellar ages, as discussed in Section~\ref{lickmeasurements}. Due to the variety of the redshifts in the sample, at all redshifts the stacked spectra overlap approximately in the wavelength range $4,150<$\AA{}$<4,800$ (shaded yellow area in Figure~\ref{fig: stacked}).

We fit our stacked spectra with \texttt{pPXF} \citep{cappellari2004}, adopting  stellar templates from version 9.1 of the MILES library, which consists of 985 stars, whose spectra  cover a range of 3,525--7,500 \AA{} at 2.51 \AA{} (FWHM) spectral resolution \citep{sanchez2006,falconbarroso2011}. During the \texttt{pPXF} fit, we use a $4^{th}$-order additive and $4^{th}$-order multiplicative polynomial correction for the spectral slope of our template. Additive polynomials are introduced to correct low-frequency differences in shape between the galaxy and the templates \citep{cappellari2004}. The multiplicative polynomials are included to ensure that the results are insensitive to the normalization or flux calibration of galaxy and stellar template spectra \citep{Kelson2000}. The polynomial degree is chosen to maximize the quality of the fit, which is also confirmed by a visual inspection. 

We find values for the velocity dispersions of between 125 and 215 km s$^{-1}$, after subtracting the instrumental velocity dispersion in quadrature. We adopt the same procedure as \cite{cappellari2009} and compute $\sigma_{\mathrm{stack}}$ using $\sigma_{\mathrm{stack}}\approx c\Delta z/(1 + z)$, where $c$ is the speed of light and $\Delta z=110\mathrm{\,km\,s}^{-1}$ is the zCOSMOS-bright redshift error. We find values for $\sigma_{\mathrm{stack}}\,\mathrm{of}\sim60-90\mathrm{\,km\,s^{-1}}$, depending on the redshift. The best-fit templates are shown as dashed red lines in Figure~\ref{fig: stacked} and Figure~\ref{fig: allspectra}. 

\begin{figure*}
\centering
\includegraphics[width=158mm]{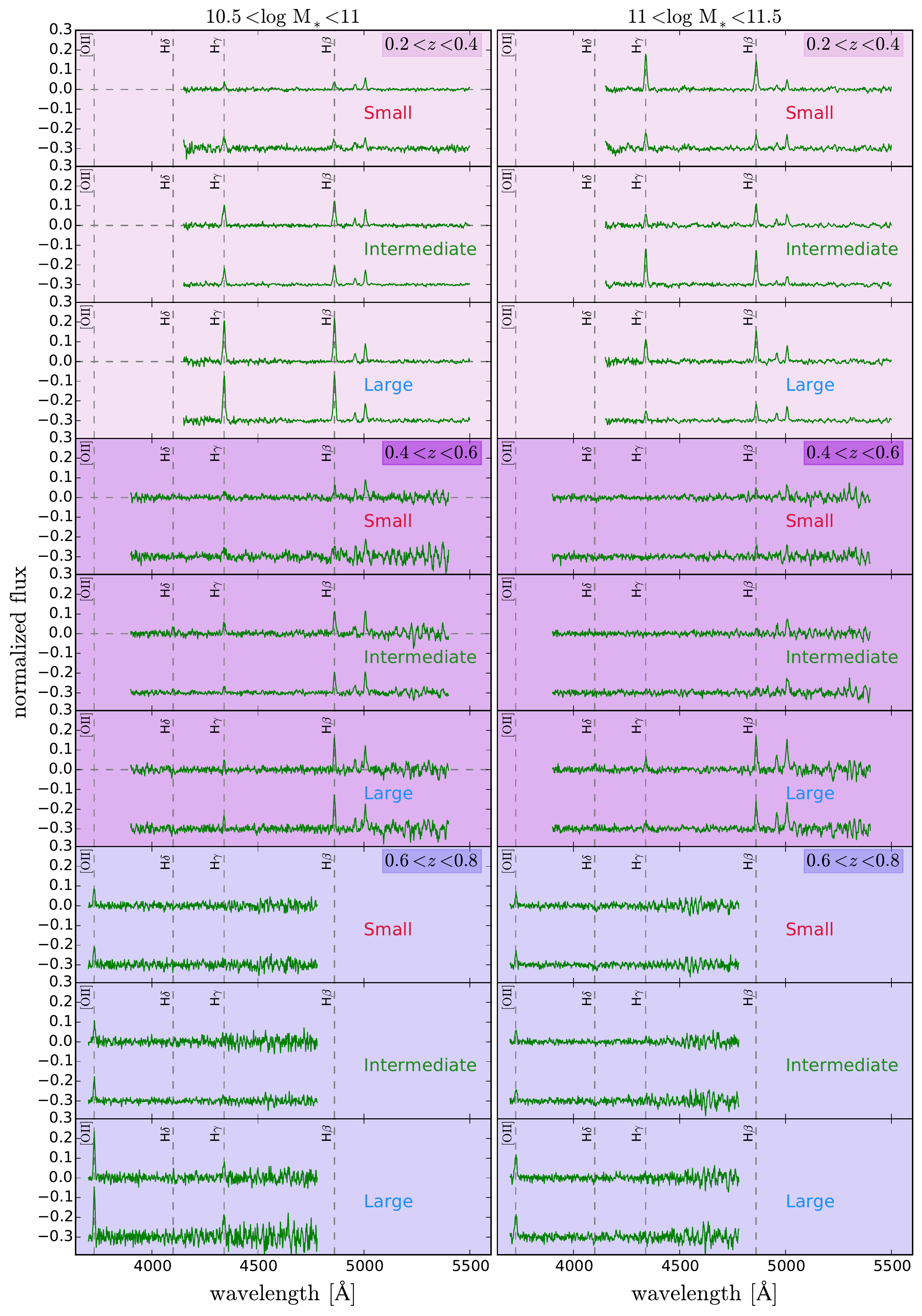}
\caption{Residuals between the observed stacked spectra and the best-fit templates from the previous figure. The upper residuals come from the size-mass cut and the lower ones from the horizontal cut, having been shifted by a constant. For less massive galaxies, the residuals in the regions of H$\beta$, H$\gamma$, H$\delta$ and [OII] appear to show higher emission line contamination with increasing sizes. The same is not however seen for more massive galaxies. This effect is visible for galaxies binned with both a size-mass and an horizontal cut. However, from fitting the residuals we obtain very low sSFRs ($\log\mathrm{(sSFR/Gyr}^{-1}) < -2$)}
\label{fig: allres}
\end{figure*}

For each of the stacked spectra of Figure \ref{fig: allspectra}, we show greatly expanded in Figure~\ref{fig: allres} the residuals obtained after subtraction of the best-fit stellar-superposition templates.  The regions around the Balmer lines, [OIII] and [OII] have been masked with the \texttt{goodpixels} function provided in \texttt{pPXF}. For the small galaxies, the residual spectra are also shown along the bottom of each panel in Figure \ref{fig: stacked}.

In each panel of Figure \ref{fig: allspectra}, the upper spectrum refers to the cut along the size-mass relation, the lower one to  the horizontal cut. The horizontal cut spectra have been shifted by subtracting a constant value from the normalized spectra. This is also done for the residuals in Figure~\ref{fig: allres}. The different background colors in both figures indicate different redshift bins.   

It is clear that all of the stacked spectra are dominated by the typical features of an old stellar population, such as a strong 4000 \AA{} break, the G-band at 4300 \AA{}, the 5180 \AA{} Mg and the Balmer absorption features.

Nevertheless, it is clear that the residuals from essentially all of the stacks show a weak narrow emission line contamination. For masses $10.5<\log\,\mathrm{M/M}_\odot<11$ (first column of residuals in Figure~\ref{fig: allres}), the emission line fill-in appears to be stronger for larger galaxies, especially in the H$\beta$, H$\gamma$ and [OII] region. The same does not seem to apply to more massive galaxies (second column of residuals in Figure~\ref{fig: allres}). 
We fit the residual emission lines and compute the equivalent width of [OII] and H$\beta$. From the relation between sSFR, EW(H$\alpha$) and EW([OII]) we obtain $\log\mathrm{(sSFR/Gyr}^{-1}) < -2$ (see for example \citealt{moresco2013} for the relation between these quantities in zCOSMOS galaxies). We assume in deriving this the Balmer line ratios for the case B recombination with a temperature of 1$0^4$ K and a typical electron density $\leq\,10^4$ cm$^{-3}$ without reddening \citep{osterbrock2006} for the conversion between H$\alpha$ and H$\beta$ fluxes.   This is a very low sSFR,  $<10^{-11}$ yr$^{-1}$; this corresponds closely to the inverse age of the universe at $z\sim0.3$.

\subsection{Emission Lines Correction}\label{emcorrection}

We correct for the emission line contribution that is visible in Figure~\ref{fig: allres} by fitting a Gaussian to the lines in the residual spectrum, with fitting at the same time the residual continuum level, which is very close to zero.  We find velocity dispersions for the residual emission lines of H$\beta$, H$\gamma$ and H$\delta$ lines that are consistent with the instrumental resolution. We then subtract the fitted emission lines from the observed spectra. In some bins, especially in the highest redshift one, the correction is not applied, as the residuals corresponding to the Balmer lines are barely distinguishable from the noise. In order to check whether we are performing an over-subtraction of the lines, we also fit gaussians fixing the continuum level at $1\sigma$. We find minimal or no variations between the two methods.

In the rest of our analysis, to derive stellar ages for the different stacked spectra, we use the spectra subtracted of their emission line component.

\subsection{Full Spectral Fitting}\label{fullspectralfitting}

We first estimate ages by using \texttt{pPXF} for a full spectral fitting analysis, using a set of SSP templates with solar metallicity and Salpeter IMF from BC03 and exclude templates with age older than the age of the universe at each redshift and younger than 1 Gyr. The metallicity of early-type galaxies at these masses is expected to be solar with a scatter of 0.1 dex (see also \citealp{gallazzi2005, gallazzi2014, thomas2010, conroy2012}). During the \texttt{pPXF} fit we use a 10th order multiplicative polynomial correction and no additive polynomials. We use the emission line subtracted spectra as described in Section~\ref{emcorrection}, without masking the regions of Balmer lines but masking out the regions corresponding to [OIII] and [OII]. We compute the mass-weighted age among the best-fit templates found from \texttt{pPXF}.

\begin{figure}
\centering
\includegraphics[height = 91mm]{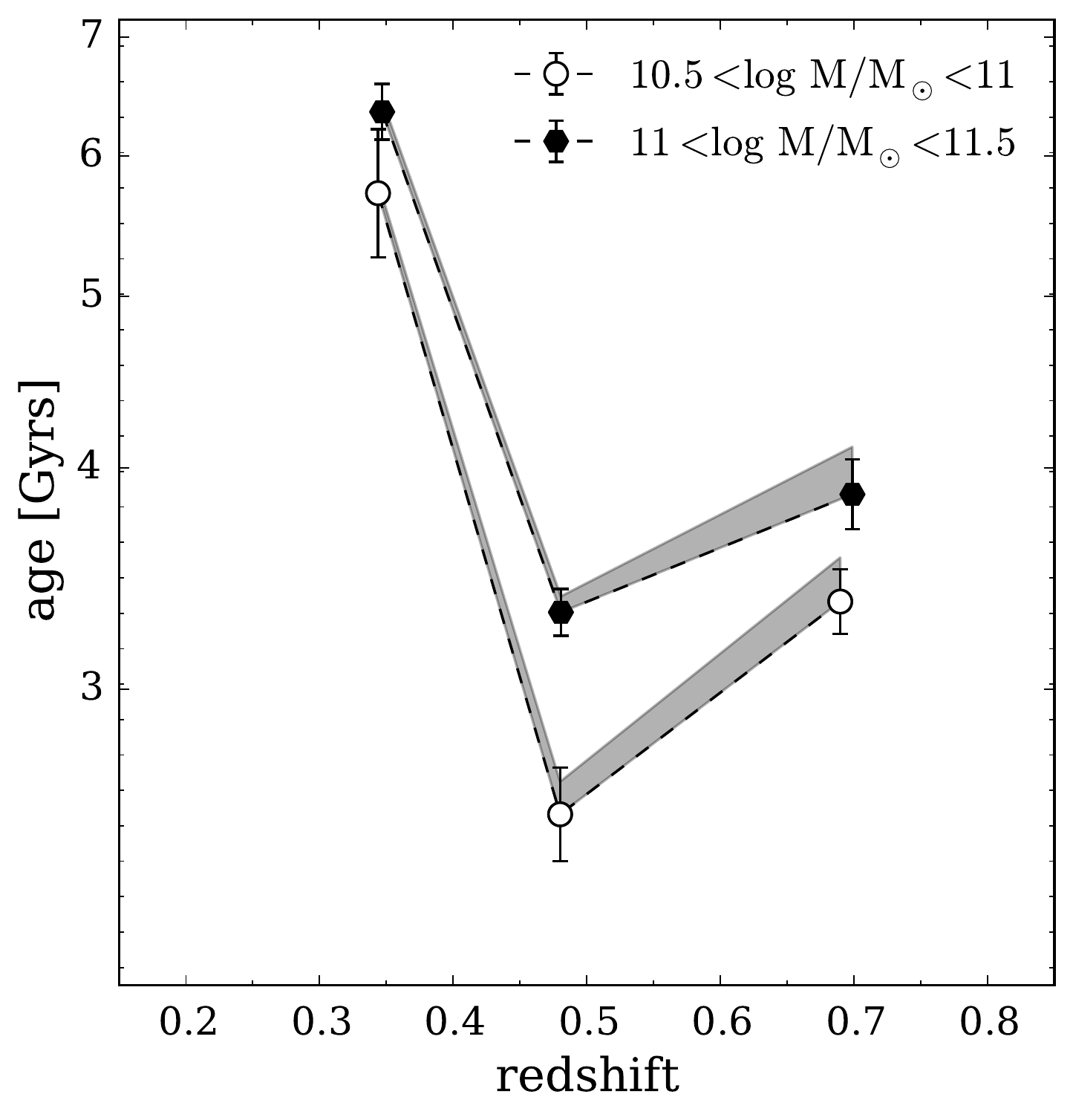}
\caption{Ages of passive galaxies derived from stacked spectra with the full spectra fitting as function of redshift and mass, with no splitting for size. More massive galaxies (filled black points) show older ages than less massive ones (empty circles) at any given redshift. The shaded areas represents the difference in ages that are derived with applying the correction for emission lines fill-in from the continuum level or from $1\sigma$ above the continuum.}
\label{fig: age_tot} 
\end{figure}

To derive the error (error bars in Figures~\ref{fig: age_tot} and~\ref{fig: results_age}) on our age estimation, we use the \textit{jackknife} technique:

\begin{equation}
\sigma_{full}^2 = \frac{N-1}{N}\sum_{i=1}^{N} (\mathrm{Age_{full}}-\mathrm{Age_{full}}_{(i)})^2
\end{equation}
where N is the number of objects stacked in each bin, $\mathrm{Age_{full}}$ is the age measured on all N spectra, and $\mathrm{Age_{full}}_{(i)}$ is the age measured on a stacked spectrum made of $N-1$ spectra by removing the $i^{\rm th}$ spectrum.
The ages obtained with this method are shown in Figures~\ref{fig: age_tot} and~\ref{fig: results_age} and listed in Table~\ref{fig: allvalues}.

\subsection{Lick Indices Analysis}\label{lickmeasurements}

To derive the strengths of the spectral absorption features we use the bandpasses and pseudo-continua for each index from \cite{trager1998}.  The resolution of the 20 indices that are covered in our different redshift ranges varies from 10.9 \AA{} ($\sim340$ km/s) for H$\delta_\mathrm{A}$ to 8.4 \AA{} ($\sim200$ km/s) for Fe5406 (see Table~\ref{indexused} for the indices  we use at each redshift bin).  The zCOSMOS-bright spectra have a resolution that is better than the defining Lick spectra for some wavelengths and worse for others.  To correct the  measured values from our spectra to the Lick resolution, we carry out the following procedure: 

\begin{equation}
I_{\mathrm{corr}}\,=\,I_{\mathrm{obs}}\,\frac{I_{\mathrm{best}}^{\mathrm{Lick}}}{I_{\mathrm{best}}^{\mathrm{obs}}},
\end{equation} 
where $I_{\mathrm{best}}^{\mathrm{obs}}$ is the index derived from the  \texttt{pPXF} best-fit at the observed velocity dispersion and $I_{\mathrm{best}}^{\mathrm{Lick}}$ on \texttt{pPXF} original template, convolved to the Lick resolution. The best-fit template we use are those derived in Section~\ref{stacking} with MILES stellar library. For all indices the  complete set of Lick's resolutions is from \cite{schiavon2007}. We derive the stellar population parameters from the corrected indices $I_{\mathrm{corr}}$. In order to get the correct Lick resolution for $I_{\mathrm{best}}^{\mathrm{Lick}}$, the template FWHM = 2.51 \AA{} has been taken into account. 

In order to check whether the index values depend on the choice of the templates, we derived, for those cases where the zCOSMOS-bright resolution was higher than the original Lick resolution, stellar population parameters from the spectra directly convolved down to the Lick resolution, without using the best-fit templates.  For these test cases, there are no significant variations from the values reported here. To derive errors on the indices, we use the \textit{jackknife} technique as explained in Section~\ref{fullspectralfitting}.

\begin{table}\label{fig: indexused}
\caption{Indices available in each redshift bin, highlighting  those  used to derive the stellar population parameters.}

\centering
\begin{tabular}[t]{p{1.8cm}p{6.2cm}}

Redshift & \\
\hline
  &  \\
$0.2 < z < 0.4$ & \hl{$\mathrm{Ca4227}$}, \hl{$\mathrm{G4300}$}, \hl{H$\gamma_\mathrm{A}$}, \hl{H$\gamma_\mathrm{F}$}, \hl{$\mathrm{Fe4383}$}, $\mathrm{Ca}4455$, \hl{$\mathrm{Fe4531}$}, $\mathrm{C}_24668$, $\mathrm{H}\beta$, $\mathrm{Fe}5015$, \hl{$\mathrm{Mg}_1$}, \hl{$\mathrm{Mg}_2$, $\mathrm{Mg}b$}, \hl{$\mathrm{Fe5270}$}, \hl{$\mathrm{Fe5335}$}, \hl{$\mathrm{Fe5406}$}\\ 
   &   \\
$0.4 < z < 0.6$ & \hl{H$\delta_\mathrm{A}$}, H$\delta_\mathrm{F}$, $\mathrm{CN}_1$, $\mathrm{CN}_2$, \hl{$\mathrm{Ca4227}$}, \hl{$\mathrm{G4300}$}, \hl{H$\gamma_\mathrm{A}$}, \hl{H$\gamma_\mathrm{F}$}, \hl{$\mathrm{Fe4383}$}, $\mathrm{Ca}4455$, \hl{$\mathrm{Fe4531}$}, $\mathrm{C}_24668$, $\mathrm{H}\beta$, $\mathrm{Fe}5015$, \hl{$\mathrm{Mg}_1$}, \hl{$\mathrm{Mg}_2$}, \hl{$\mathrm{Mg}b$}, \hl{$\mathrm{Fe5270}$}, \hl{$\mathrm{Fe5335}$}\\
  &  \\
$0.6 < z < 0.8$ & \hl{H$\delta_\mathrm{A}$}, H$\delta_\mathrm{F}$, $\mathrm{CN}_1$, $\mathrm{CN}_2$, \hl{$\mathrm{Ca4227}$}, \hl{$\mathrm{G4300}$}, \hl{H$\gamma_\mathrm{A}$}, \hl{H$\gamma_\mathrm{F}$}, \hl{$\mathrm{Fe4383}$}, $\mathrm{Ca}4455$, \hl{$\mathrm{Fe4531}$}, $\mathrm{C}_24668$\\
  &  \\

\hline
\end{tabular}
\label{indexused}
\end{table}

We derive stellar population parameters as in \cite{onodera2014}, comparing our measured indices with those computed on SSP models with variable $\alpha$-abundances \citep{thomas2011}. The code makes a 3D grid with an uniform interval of 0.02 dex of the \cite{thomas2011} model values for age, [Z/H], and [$\alpha$/Fe]. The model spans the parameter space $0.1 < \mathrm{age/Gyr} < 15$, $-2.25 < [\mathrm{Z/H}] < 0.67$, and $-0.3 < $[$\alpha$/Fe]$ < 0.5$ and adopts the Salpeter IMF. The best-fit values of these three stellar population parameters are derived by comparing our corrected indices with the \cite{thomas2011} model and finding a set of parameters which gives minimum $\chi^2$:

\begin{equation}
\chi^2\,=\sum\,\frac{(I_{\mathrm{Thomas}}\,-\,I_{\mathrm{measured}})^2}{\sigma^2_{\mathrm{Lick}}}
\end{equation}
where $\sigma_{\mathrm{Lick}}$ is the error bar on our Lick indices computed with the \textit{jackknife} procedure. This allows us to make use of all available indices at the same time and returns best-fit values for all the three free parameters. We use all the indices available in each redshift bin, with few exceptions.  
The indices Ca4455, H$\delta_{\mathrm{F}}$, Fe5015 and H$\beta$ have been excluded following \cite{thomas2011} recommendation. The iron line Fe5015 is a problem due to its proximity to the strongest of the emission lines of the doublet [OIII] at 5007 \AA. The Balmer line H$\beta$ is also known to be problematic in deriving ages, as it is affected by fill-in from the emission line more than higher order Balmer lines  (see also \citealp{poole2010, onodera2014}). Ca4455 is found to be mostly sensitive to Ca abundance \citep{korn2005, thomas2011} and its use should be therefore taken with caution, as it is also demonstrated from comparison with globular clusters \citep{korn2005}. We also excluded other indices sensitive to the C abundance, namely CN$_1$, CN$_2$ and C$_{2}$4668, as the available models do not use the C abundance as a free parameter (see also \citealt{thomas2011, onodera2014}).

We used Ca4227, G4300, H$\gamma_\mathrm{A}$, H$\gamma_\mathrm{F}$, Fe4383, Fe4531 for all redshifts, as these indices lie in the common wavelength range highlighted in yellow in Figure~\ref{fig: stacked}. The Balmer lines are the most sensitive indices to age variations. The index Fe4383 is particularly sensitive to the iron abundance and it is also influenced by total metallicity [Z/H] and by the magnesium abundance. Fe4531 and Ca4227 are also sensitive to the total metallicity. Ca4227 is also dominated by variations in Ca and C abundances and show very weak dependence to $\alpha$ variations. The same can be said about the G4300 band, whose main contributors are C, O and Fe abundances. We included H$\delta_{\mathrm{A}}$, Mg$_1$, Mg$_2$, Mg$b$, Fe5270, Fe5335 and Fe5406, when available. H$\delta_{\mathrm{A}}$ is sensitive to age variations, despite being more affected by metalliticy than H$\gamma_{\mathrm{A}}$ and H$\beta$, and is also important for [$\alpha$/Fe] ratio estimation, as the higher order Balmer indices like H$\delta_{\mathrm{A}}$ are sensitive to $\alpha$ enhancement \citep{thomas2004}.  Mg$_1$, Mg$_2$ and Mg$b$ are the major indicators of the $\alpha$-element abundance, although being affected also by [Z/H], C and Fe. The Mg$_2$ band-pass covers also the Mg$b$ lines, the sensitivity of this index is therefore higher than Mg$_1$. Fe5270, Fe5335 and Fe5406 show similar trends and are sensitive mostly to Fe, [Z/H] and Mg abundances. Fe5406 is only available at low redshift ($0.2<z<0.4$). Its strength is however weaker than the other two and its inclusion does not modify the results.

In deriving stellar population parameters we first leave [Z/H] as a free parameter. However,  the results are strongly affected by the spurious age--metallicity anticorrelation mentioned in the Introduction, especially in the redshift bin $0.4<z<0.6$. In an alternative approach, we restrict the [Z/H] values to those from the mass--metallicity relation from \cite{gallazzi2005} obtained in the local Universe but evidently applicable also up to  $z=0.7$ because of the apparent lack of evolution in this redshift range \citep{gallazzi2014}. We correct their masses to our integrated SFRs with a correction factor of $\sim0.2$ dex.

We derive ages using both the indices from  the original stacked spectra as well as those from spectra corrected for the emission lines as described in Section~\ref{emcorrection}. As expected, on average slightly younger ages are obtained from the emission-corrected Balmer indices, with a stronger correction for large galaxies in the less massive bin (see residuals in Figure~\ref{fig: allres}). The effect of the age-metallicity degeneracy are much less severe in the emission line corrected spectra than in the uncorrected ones, though still visible. This is most likely due to the fact that it was more difficult to find a unique solution for age and [Z/H] when the uncorrected Balmer indices were forcing  older ages.
The results we obtain with the Lick system are listed in Table~\ref{fig: allvalues}.

\begin{figure*}
\centering
\includegraphics[height = 95mm]{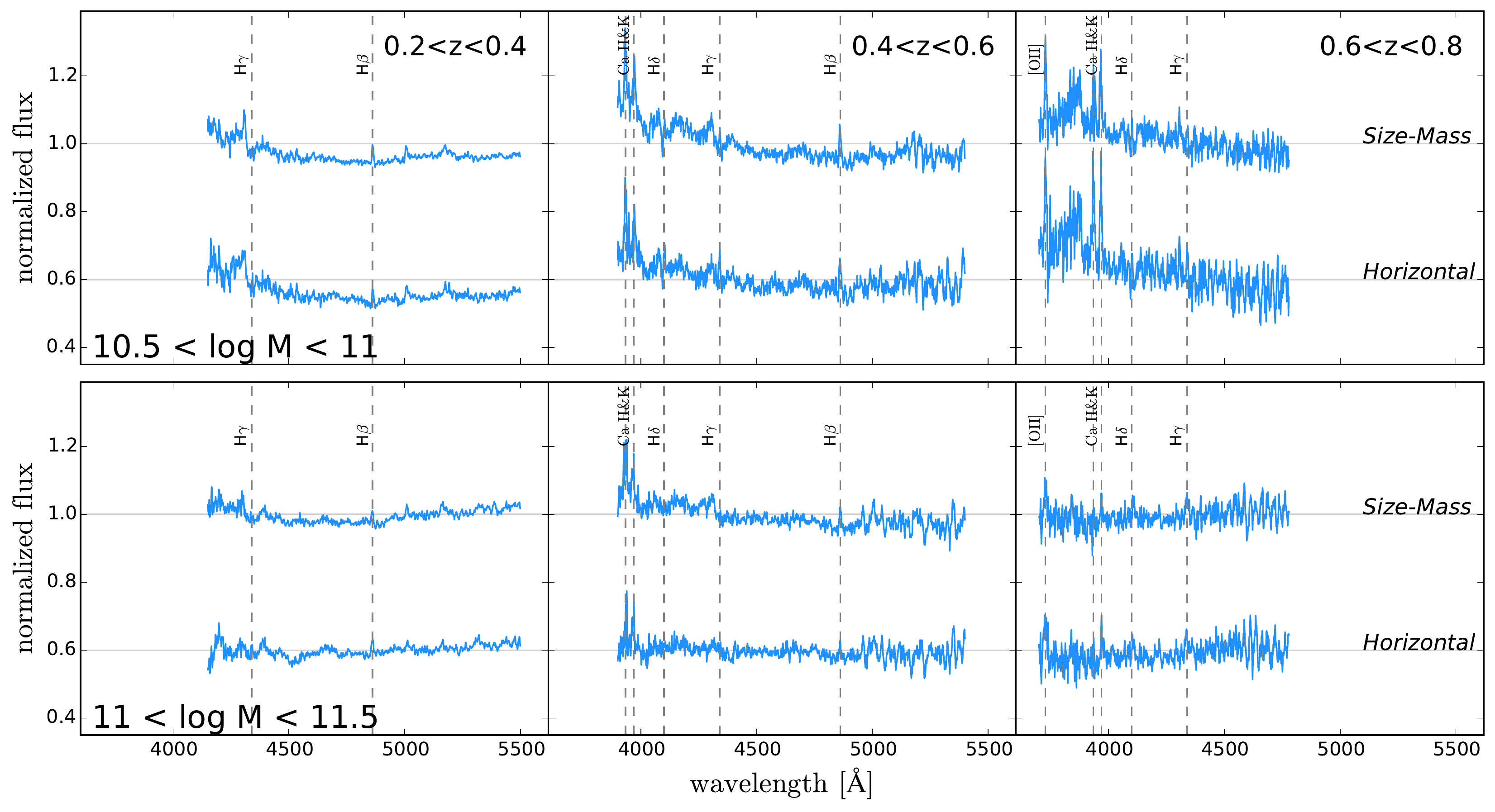}
\caption{Ratio between the stacked spectra of large and small galaxies for the size-mass cut (top spectrum in each panel), and for the horizontal cut (bottom spectrum), the latter vertically shifted by subtracting a constant value.  At the lower masses, the overall redness of the small spectrum vs. the large is clearly visible, as well as the much deeper Calcium H and K lines and the suppression of flux shortward of 4000 \AA{} in the small spectrum. These features are both expected for older ages. The situation with the Balmer lines is less clear as stronger absorption would be expected in the younger object. This effect could however be greatly weakened because of the effects of residual line emission, as seen in Figure~\ref{fig: allres}. Interestingly, the same cannot be said about the higher mass spectra, with no hint of deeper H and K and stronger 4000 \AA{} break. Similar behaviour can be identified for galaxies stacked with both the horizontal cut and the cut along the size-mass relation.}
\label{fig: ratio_large_small} 
\end{figure*}

 \begin{figure*}
\centering
\includegraphics[height = 135mm]{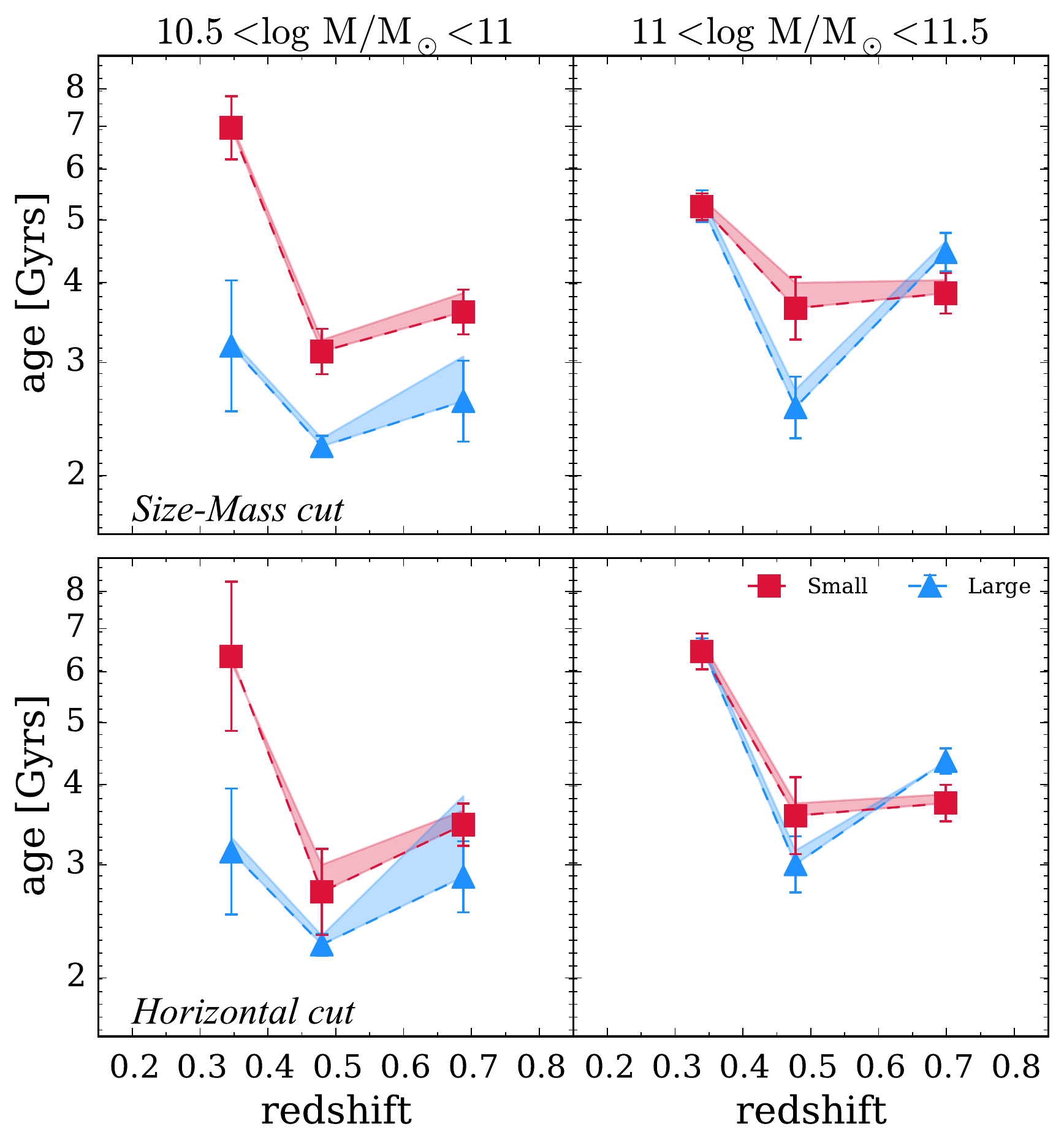}
\caption{Age results for our sample, obtained with the full spectral fitting technique. Two different panels show the two different mass bins. At masses $10.5<\log\mathrm{M/M_\odot<11}$ (left panel) red symbols (small galaxies) are systematically the oldest in each redshift bin, while the blue points (largest galaxies) are the youngest objects at each redshift. The difference apparently increases with cosmic time. At higher masses $11<\log\mathrm{M/M_\odot<11.5}$, the behavior is less clear-cut. There is no clear and significant difference in the evolution among different sizes. The trends for the low mass bin remain similar with both the size-mass cut (top panels) and horizontal cut (bottom panels). The shaded areas represents the difference in ages derived with applying the correction for emission lines fill-in from the continuum level or from $1\sigma$ above the continuum.}
\label{fig: results_age} 
\end{figure*}

\begin{figure*}
\centering
\includegraphics[height = 114mm]{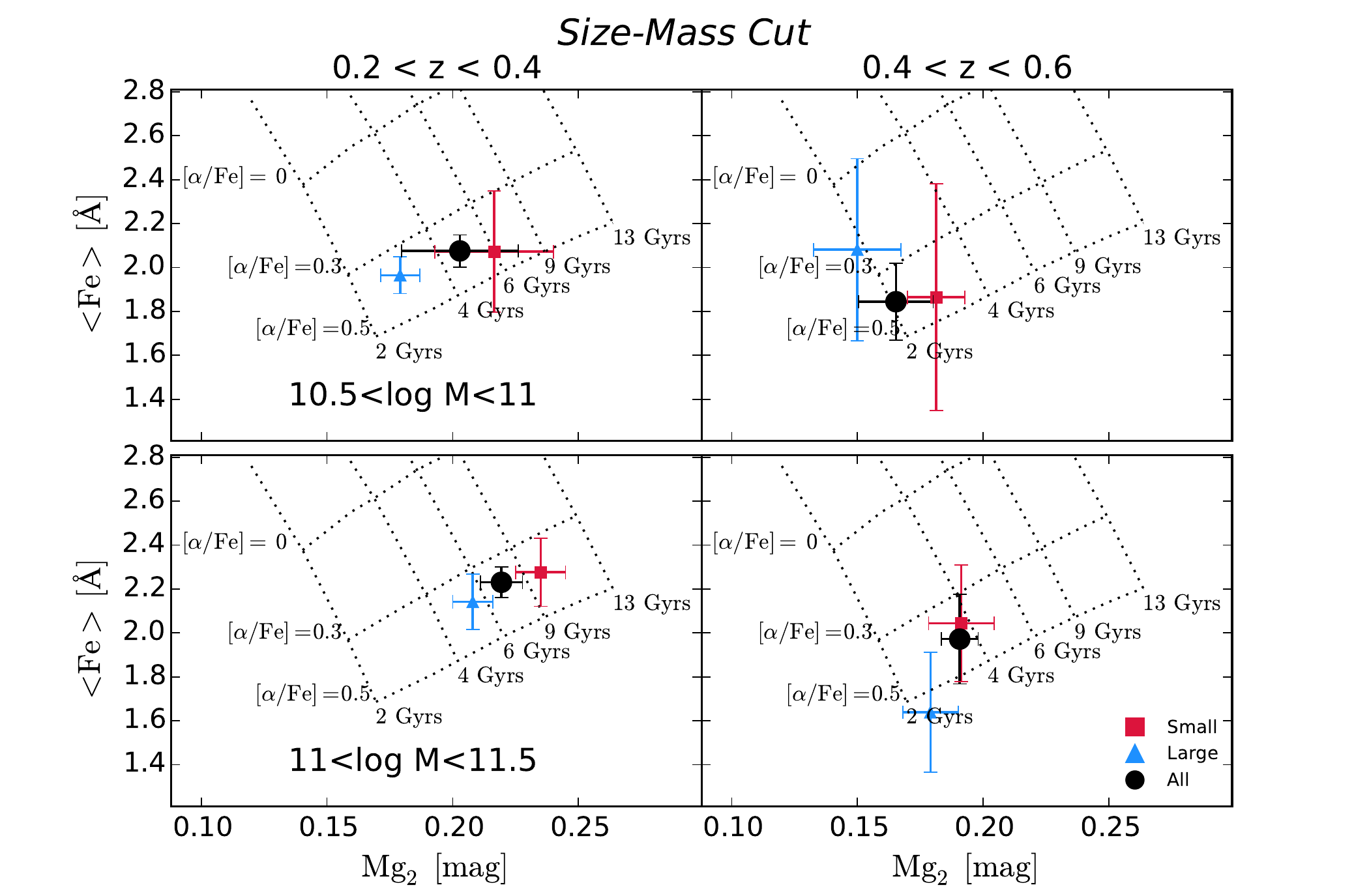}
\caption{$<\mathrm{Fe}>$ vs Mg$_2$, where $<\mathrm{Fe}>$ = (Fe5270+Fe5335)/2 from the Lick indices. These are the values obtained from the size-mass cut. Model lines are from \cite{thomas2011}. Dotted lines represent loci of constant [$\alpha$/Fe] and constant age (both at solar metallicity). The upper panels show the results for the lower mass bin, the lower panels for the higher mass bin. We show only the lowest two redshift bins as at high-$z$ we do not have wavelength coverage for Mg, Fe5270 and Fe5335.  We do not find any significant difference in $\alpha$ enhancement for galaxies at different masses and redshifts.}

\label{fig: mgfe} 
\end{figure*}

\begin{figure*}
\centering
\includegraphics[height = 114mm]{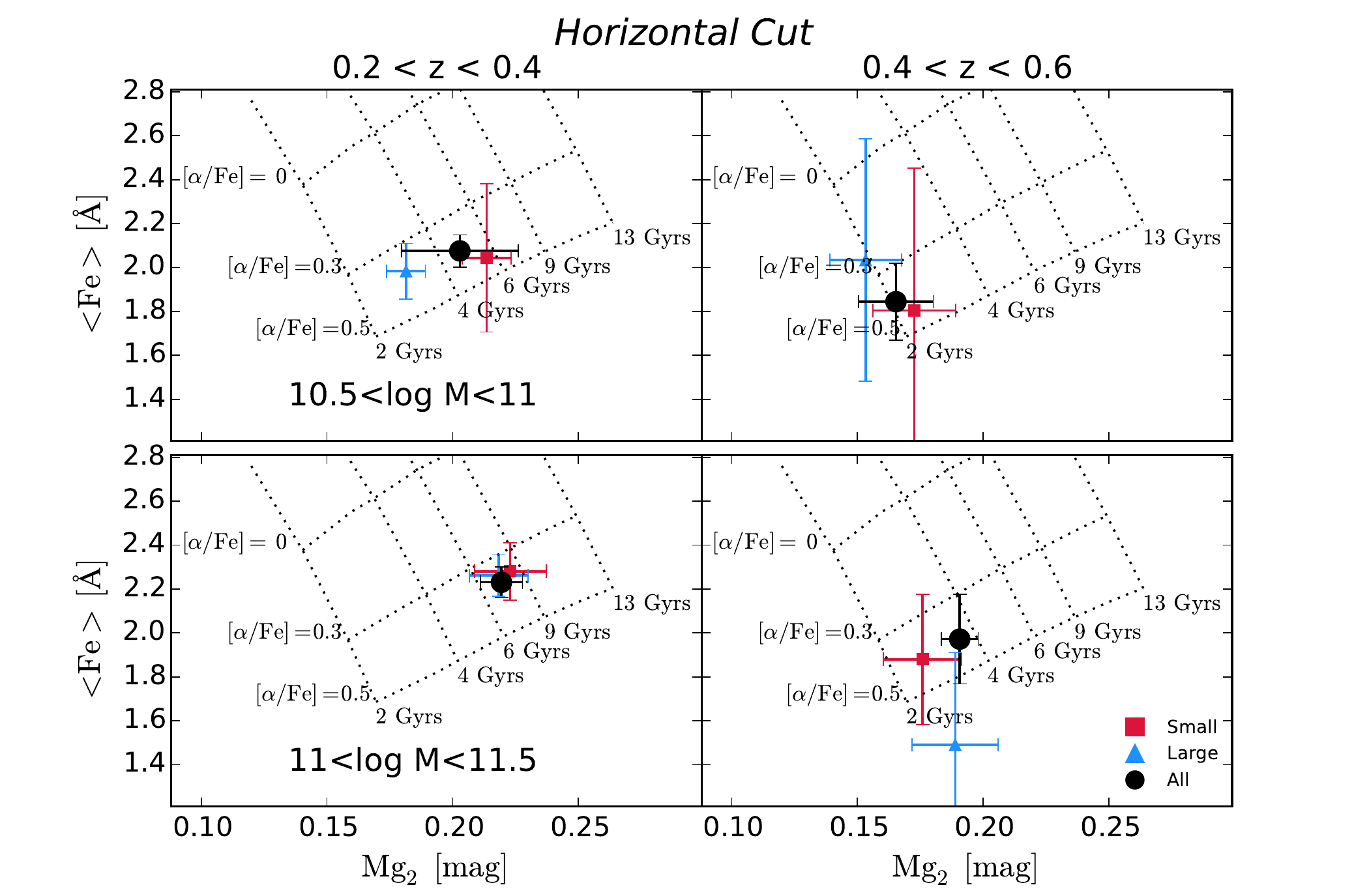}
\caption{The same as Figure~\ref{fig: mgfe}, but for the horizontal cut in sizes. Again, we do not find differences in $\alpha$-enhancement for galaxies of different masses or redshifts.}
\label{fig: mgfe_horizontal} 
\end{figure*}

 \section{Results}\label{results}

Before examining the age difference between large and small galaxies, we look first at the ages of passive galaxies, at each redshift, as a function of their stellar mass, but independent of their sizes. \cite{thomas2010} have shown that, at least at redshift zero, there is a clear mass-age relation for quenched galaxies, in the sense that more massive quenched galaxies are found to be on average older than their less massive counterparts.
 
In Figure~\ref{fig: age_tot} we show the stellar population ages derived from the stacked spectra of our quenched galaxies, in bins of stellar mass and redshift, using in each mass-redshift bin all the galaxies, regardless of size. More massive galaxies in our sample (filled black circles) indeed have older ages than less massive ones (empty circles) at any given redshift, in agreement with the previous studies at $z\sim0$, but now extending this result out to $z=0.8$.  The shaded areas represents the difference in ages obtained when applying the correction for emission lines fill-in from the continuum level or from $1\sigma$ above the continuum. The decrease in age seen in the redshift bin $0.4<z<0.6$ might conceivably be related to the fact that these galaxies lie in a less dense region of the universe with respect to the known over-densities of COSMOS at 0.3 \citep{knobel2009} and 0.7 \citep{guzzo2007, scoville2007}. The over-dense regions are also visible in Figure~\ref{fig: measurements}.

Overall this test, reproducing the well-known mass-age relation for quenched galaxies and extending this at $z>0$ redshifts,  gives us confidence that our measurements of ages, at least in a comparative sense, are fairly robust. We therefore turn to the main question of our paper, which is to establish the age comparison between quenched galaxies of different sizes at constant redshift and stellar mass.

\subsection{The stellar ages of small and large quenched galaxies}

\subsubsection{Visual Inspection of spectral ratios}\label{visualinspection}
 
Before proceeding to a more quantitative analysis, significant differences between the spectra of large and small galaxies are readily visible `by eye'. In Figure~\ref{fig: ratio_large_small} we show  the ratio of the spectra of large/small galaxies. At the lower masses, the overall redness of the small spectrum vs. the large one is clearly visible, as well as the much deeper Calcium H and K lines and the suppression of flux shortward of 4000 \AA{} in the small spectrum. These features are both expected for older ages. The situation with the Balmer lines is less clear cut, as stronger absorption would be expected in the younger object, but this could be weakened by the residual emission which can be seen in Figure~\ref{fig: allres}. 
Interestingly, the same differences are not seen in the higher mass stacked spectra. In the ratio of large to small galaxies for more massive galaxies (bottom panels in Figure~\ref{fig: ratio_large_small}), deeper Ca H$\&$K lines for smaller galaxies can still be seen in the redshift bin $0.4<z<0.6$. On the other hand, no hint of deeper H and K and stronger 4000 \AA{} break is visible at higher $z$, showing therefore a less clear trend of age with size than in less massive objects.
Similar behavior is seen at all redshifts for the spectra whether they are stacked with the horizontal cuts or with the cuts along the size-mass relation.

\subsubsection{Ages of small and large quenched galaxies: Quantitative estimates from the full spectral fitting and Lick indices approaches}

In Figures~\ref{fig: results_age} we show our best-fit stellar population ages as a function of redshift for the large and small galaxies, obtained with the full spectral fitting technique using \texttt{pPXF}. The spectra have been corrected for emission line contribution visible in the residuals in Figure~\ref{fig: allres} and fitted with a combination of SSP with solar metallicity. The results show a clear trend in ages for less massive galaxies. 

At masses $10.5<\log\mathrm{M_*/M_\odot<11}$ (left panel) red symbols (small galaxies) are the oldest in each redshift bin, while blue points (large galaxies) are the youngest objects at each redshift, with the age difference between small and large galaxies being especially evident in the lowest redshift bin. Same trends are visible for both the size-mass and the horizontal cuts.   The Lick-based analysis that was described in Section~\ref{lickmeasurements} also shows the same trends, though overall younger ages are returned for both the small and large galaxies, as expected since the model from \cite{thomas2011} returns luminosity-weighted ages.  However, the Lick-derived ages reported in Table~\ref{fig: allvalues}, which were computed with metallicity as a free parameter, should be treated with some caution because of the strong age-metallicity degeneracy.

For masses $11<\log\mathrm{M_*/M_\odot<11.5}$, the behavior is different and a clear difference between the ages of galaxies with different sizes is not visible.  This lack of difference is present both in the full spectral fitting or in the Lick analysis.

We stress that the results shown here are derived with using as definition of stellar masses the integral of the SFR. However, repeating the analysis using bins of stellar masses computed including the loss of mass due to the return or material to the ISM due to stellar evolution (which we have argued is not strictly correct), also gives the same qualitative results, as in the lower mass bin larger galaxies are younger than smaller one, and no clear trend with ages is visible in the higher mass bin.

In Figure~\ref{fig: mgfe} and Figure~\ref{fig: mgfe_horizontal} we show an example of a Lick index-index diagram, $<\mathrm{Fe}>$ (mean obtained with indices Fe5270 and Fe5335) versus Mg$_2$ (see for example \citealt{burstein1984, carollo1994, fisher1996}), for the size-mass and the horizontal cut, respectively. We show results only for the two lowest redshift bins, as at the highest redshift we do not have the wavelength coverage for Mg$_2$, Fe5270 and Fe5335. The model lines are from \cite{thomas2011}. Dotted lines show loci of constant age (2, 4, 6 and 9 Gyrs) and constant [$\alpha$/ Fe], all computed at fixed solar metallicity. Small galaxies (red squared points) and large galaxies (blue triangles), while clearly showing differences in ages, do not show differences in [$\alpha$/Fe] for different sizes. The black points show the index values for galaxies with same mass and redshift binning as the others in the same plot, but with no split for sizes.  All the galaxies appear to have super-solar $\alpha$ to Iron ratios ($[\alpha$/Fe]$>0.3$), as expected for massive early-type galaxies (e.g., \citealt{leeworthey2005, thomas2005, choi2014}). 

\begin{table*}
\caption{The ages derived from full spectral fitting (first column) and  Lick indices, with fixed [Z/H] (second column) and free [Z/H] (third column). The ages from the Lick indices have been in both cases derived from emission lines free spectra. The number in parenthesis show the results obtained by fitting the residual emission lines having as base level $1\sigma$ above the continuum.}\label{fig: allvalues}
\centering
\begin{tabular}{|llccc|ccc|}
\hline
& & & & & & &\\

{Size-Mass cut} & &  &  $10.5<\log\mathrm{M<11}$ & &  & $11<\log\mathrm{M<11.5}$ &\\ 
\hline

{ } & { } & {Spectral Fitting} & {Lick (fixed [Z/H])} & {Lick (free [Z/H])} & {Spectral Fitting} & {Lick (fixed [Z/H])} & {Lick (free [Z/H])}\\ 
{Redshift} & {Size} & {Log (Age/Gyrs)} & {Log (Age/Gyrs)} & {Log (Age/Gyrs)} & {Log (Age/Gyrs)} & {Log (Age/Gyrs)} & {Log (Age/Gyrs)}\\ 
\hline

{ $0.2<z<0.4$} & {Small} & {$0.84(0.85)^{+ 0.05}_{-0.05}$}& {$0.64(0.68)^{+ 0.04}_{-0.02}$}& {$0.70(0.74)^{+ 0.08}_{-0.08}$}& {$0.72(0.73)^{+ 0.02}_{-0.02}$}& {$0.44(0.46)^{+ 0.02}_{-0.02}$}& {$0.12(0.12)^{+ 0.02}_{-0.02}$}\\
{  } & {Large} & {$0.50(0.51)^{+ 0.10}_{-0.10}$}& {$0.28(0.30)^{+ 0.04}_{-0.04}$}& {$0.02(0.04)^{+ 0.06}_{-0.04}$}& {$0.72(0.74)^{+ 0.02}_{-0.02}$}& {$0.42(0.44)^{+ 0.04}_{-0.02}$}& {$0.20(0.22)^{+ 0.08}_{-0.06}$}\\
{ $0.4<z<0.6$} & {Small} & {$0.49(0.51)^{+ 0.04}_{-0.04}$}& {$0.42(0.44)^{+ 0.02}_{-0.02}$}& {$0.40(0.46)^{+ 0.10}_{-0.06}$}& {$0.56(0.60)^{+ 0.05}_{-0.05}$}& {$0.48(0.50)^{+ 0.02}_{-0.04}$}& {$0.60(0.60)^{+ 0.12}_{-0.10}$}\\
{  } & {Large} & {$0.35(0.36)^{+ 0.02}_{-0.02}$}& {$0.30(0.32)^{+ 0.02}_{-0.02}$}& {$0.30(0.32)^{+ 0.10}_{-0.08}$}& {$0.41(0.44)^{+ 0.05}_{-0.05}$}& {$0.30(0.34)^{+ 0.02}_{-0.02}$}& {$0.32(0.36)^{+ 0.04}_{-0.06}$}\\
{ $0.6<z<0.8$} & {Small} & {$0.56(0.59)^{+ 0.03}_{-0.03}$}& {$0.36(0.36)^{+ 0.02}_{-0.02}$}& {$0.44(0.44)^{+ 0.18}_{-0.14}$}& {$0.59(0.61)^{+ 0.03}_{-0.03}$}& {$0.26(0.26)^{+ 0.04}_{-0.04}$}& {$0.34(0.34)^{+ 0.28}_{-0.22}$}\\
{  } & {Large} & {$0.42(0.49)^{+ 0.06}_{-0.06}$}& {$0.24(0.32)^{+ 0.02}_{-0.02}$}& {$0.18(0.28)^{+ 0.08}_{-0.08}$}& {$0.65(0.67)^{+ 0.03}_{-0.03}$}& {$0.36(0.40)^{+ 0.02}_{-0.02}$}& {$0.48(0.52)^{+ 0.16}_{-0.14}$}\\
\hline 
& & & & & & &\\
{Horizontal cut} & &  &  $10.5<\log\mathrm{M<11}$ & &  & $11<\log\mathrm{M<11.5}$&\\ 
\hline
{ } & { } & {Spectral Fitting} & {Lick (fixed [Z/H])} & {Lick (free [Z/H])} & {Spectral Fitting} & {Lick (fixed [Z/H])} & {Lick (free [Z/H])}\\ 
{Redshift} & {Size} & {Log (Age/Gyrs)} & {Log (Age/Gyrs)} & {Log (Age/Gyrs)} & {Log (Age/Gyrs)} & {Log (Age/Gyrs)} & {Log (Age/Gyrs)}\\ 
\hline

{ $0.2<z<0.4$} & {Small} & {$0.80(0.79)^{+ 0.12}_{-0.12}$}& {$0.64(0.58)^{+ 0.02}_{-0.02}$}& {$0.60(0.48)^{+ 0.06}_{-0.06}$}& {$0.81(0.83)^{+ 0.03}_{-0.03}$}& {$0.60(0.64)^{+ 0.04}_{-0.04}$}& {$0.36(0.42)^{+ 0.12}_{-0.10}$}\\
{  } & {Large} & {$0.50(0.52)^{+ 0.10}_{-0.10}$}& {$0.30(0.30)^{+ 0.04}_{-0.04}$}& {$0.04(0.04)^{+ 0.06}_{-0.04}$}& {$0.82(0.82)^{+ 0.01}_{-0.01}$}& {$0.64(0.66)^{+ 0.08}_{-0.08}$}& {$0.30(0.28)^{+ 0.20}_{-0.14}$}\\
{ $0.4<z<0.6$} & {Small} & {$0.43(0.48)^{+ 0.07}_{-0.07}$}& {$0.32(0.36)^{+ 0.02}_{-0.04}$}& {$0.36(0.44)^{+ 0.10}_{-0.08}$}& {$0.55(0.57)^{+ 0.06}_{-0.06}$}& {$0.38(0.38)^{+ 0.02}_{-0.04}$}& {$0.50(0.52)^{+ 0.14}_{-0.12}$}\\
{  } & {Large} & {$0.35(0.37)^{+ 0.02}_{-0.02}$}& {$0.30(0.32)^{+ 0.02}_{-0.02}$}& {$0.34(0.38)^{+ 0.12}_{-0.08}$}& {$0.48(0.50)^{+ 0.04}_{-0.04}$}& {$0.40(0.42)^{+ 0.02}_{-0.02}$}& {$0.44(0.50)^{+ 0.08}_{-0.06}$}\\
{ $0.6<z<0.8$} & {Small} & {$0.54(0.56)^{+ 0.03}_{-0.03}$}& {$0.32(0.34)^{+ 0.02}_{-0.04}$}& {$0.40(0.40)^{+ 0.26}_{-0.24}$}& {$0.57(0.59)^{+ 0.03}_{-0.03}$}& {$0.26(0.26)^{+ 0.04}_{-0.04}$}& {$0.70(0.42)^{+ 0.22}_{-0.30}$}\\
{  } & {Large} & {$0.46(0.58)^{+ 0.06}_{-0.06}$}& {$0.20(0.30)^{+ 0.02}_{-0.04}$}& {$0.20(0.38)^{+ 0.22}_{-0.14}$}& {$0.64(0.64)^{+ 0.02}_{-0.02}$}& {$0.36(0.38)^{+ 0.04}_{-0.04}$}& {$0.64(0.60)^{+ 0.24}_{-0.30}$}\\

\hline
\end{tabular}
\end{table*}

\section{Discussion}\label{discussion}

The analysis presented in the previous sections highlights an important change in behavior of QGs with masses respectively above and below $\sim10^{11} \mathrm{M}_\odot$: Below this mass threshold,  and precisely in our $10.5<\log\mathrm{M}_*/\mathrm{M}_\odot<11$ mass bin,   the smaller galaxies have systematically older stellar populations than their larger counterparts at each cosmic epoch. This is true for both our definitions of {\it smallness} and {\it largeness}, i.e., considering both the size-mass cut (relative sizes below the evolving size-mass relation), and the horizontal cut (absolute sizes below a constant threshold in kpc). In contrast, in our high-mass bin above 10$^{11} \mathrm{M}_\odot$, we do not detect any distinct trend for QGs of different sizes. We discuss more in detail below the implications of these different behaviors.

\subsection{The size-age relation of  $M<10^{11} M_\odot$ QGs at $0.2 \leq z \leq 0.8$}

Important implications for the origin of the observed evolution of the average size of QGs at masses $< 10^{11}\mathrm{M}_\odot$ can be inferred by putting together information on the evolution of the relative number density of smaller and larger galaxies and  the fact that, in this mass bin, the former are older than the latter.
 
In C13, the number density of compact galaxies (i.e., r$_{1/2}<2$ kpc) remains stable since $z=1$. This means that either the compact galaxies are growing in size, but some other effect is continuously adding new compact QGs (e.g., \citealt{damjanov2015a}), keeping the number density constant by replacing the galaxies that grow out of that mass bin, or that the growth of the QG population occurs through the addition of larger, newly-quenched galaxies to a stable population of compact galaxies. These scenarios differ in the sizes of the new members of the QG population. In the first case, the new galaxies are compact and the ages of the smaller galaxies should be systematically younger than those of larger galaxies at the same redshift.  In the second case, the new galaxies are larger, and the ages of the smaller galaxies would be systematically older.

The results of this paper clearly support the second scenario. Compact galaxies are \textit{older} than larger ones.  This spectroscopic result is consistent with the purely photometric trends found by C13 in the same redshift range as here, and also with similar findings  of \citet{saracco2011} at $z\sim2$ and of \citealt{vanderwel2009} at $z=0$ (see however \citet{yano2016} for opposite results obtained with photometry).

Summarizing, the younger ages of larger galaxies plus the evidence that their number density is increasing with cosmic time  together push towards the conclusion that progenitor bias is driving a large part of the apparent evolution of the median size of passive galaxies below $10^{11} \mathrm{M}_\odot$. This in itself reflects the changes in the size-mass relation of their star-forming progenitors (\citealt{toft2007, buitrago2008, kriek2009}).  As shown in \citet{lilly2016}, a half of the size difference between star-forming galaxies and QGs of a given mass is explained by progenitor bias effects in which the sizes of star-forming galaxies of a given mass increase with cosmic time, and the other half by differential radial fading after quenching of the stellar populations of the star forming progenitors (see also \citealt{tacchella2015b, carollo2016}.). About half of the average size evolution of QG emerges thus naturally by considering progenitor effects.  

Our results therefore imply that the increase with cosmic time of the number density of QGs at these `low' masses, which is well quantified in the observed mass functions of quenched galaxies at different redshifts (\citealt{williams2009, ilbert2010, ilbert2013, dominguezsanchez2011}), is mostly driven by the addition of the population of  larger, later quenched objects, with smaller galaxies keeping a stable number density though cosmic time.

This is further substantiated by the fact that, between our highest and lowest redshift bins, the $\Delta$age  of the compact galaxies (selected at constant absolute size through the horizontal cut)   is consistent with passive evolution of their stellar populations, and therefore consistent with the idea that this population has hardly added any new members.  In contrast, the smaller difference in ages for larger galaxies is consistent with the idea that new, younger and i.e., newly quenched galaxies have been added to this population.

\subsection{The size-age relation of  $M>10^{11} M_\odot$ QGs at $0.2 \leq z \leq 0.8$}

The signatures of progenitor bias are less apparent (if at all) at higher masses above $10^{11} \mathrm{M}_\odot$ (for us, specifically, in our $11<\log\mathrm{M}_*/\mathrm{M}_\odot<11.5$ mass bin).  This difference has already been pointed out by C13, and is not unexpected. There is a host of evidence that this mass represents a threshold in the importance of merging: Above and below this threshold, QGs have respectively  boxy or disky isophotes, cores or cusps, flat or  steep metallicity gradients and are slow or fast rotators (see e.g. \citealt{bender1988, carollo1993, faber2007, cappellari2013a}).  Similar results are also found in hydrodynamical simulations within a $\Lambda$CDM universe (\citealt{hopkins2009, feldmann2010}).  Furthermore,  as stressed in  \citet{peng2010},  imposing that galactic evolution obeys to a continuity equation  requires that very massive passive galaxies at $10^{11}$ M$_\odot$ must have generally undergone significant merging  after quenching (see their Figure 16).  

Also, at these very high masses, C13 report a decrease in the number density of the most compact objects between the high and low redshift bin of  $\sim30\%$, suggesting that at these masses the contribution of newly-quenched objects is less important.  Using our spectroscopic sample, we (confirm the same trends of C13 for the number density evolution of large and small QGs in the low mass bin, but)  find that the number density of  small galaxies at high masses remains stable as well, as at the low masses. Our definition of small galaxies at higher masses differs however from that in C13, as we adopt a threshold of    $<4.5$ kpc (instead of 2.5 kpc as in C13) to identify small systems. Another element for caution in handling our spectroscopically-estimated number densities comes from the possible incompleteness biases of the spectroscopic survey, although  zCOSMOS was designed to yield a high and fairly uniform sampling rate across most of the field (about 70$\%$), and has delivered  a high success rate in measuring redshifts \citep{lilly2009}.

\subsection{The timescales  for the stellar mass growth in QGs of different masses and sizes}

Our investigation of the [$\alpha$/Fe] abundance ratios for the different subsamples of QGs in our sample, i.e., QGs at different masses below and above $10^{11} \mathrm{M}_\odot$ and, within the same mass bin, of different sizes, have generally enhanced values relative to the solar value. This is a well-known property of massive QGs at redshift $z=0$ ( \citealt{carollo1993, carollo1994, leeworthey2005, thomas2005, choi2014}), and has been hinted to be true also at higher redshifts in very small samples \citep{onodera2014}.  Our study proves that, at least at masses above $10^{10.5} \mathrm{M}_\odot$, such enhanced abundance ratios, which support short timescales for the buildup  of the stellar populations of QGs,  are already achieved at least since $z=0.6$.

Furthermore, within the errors  and in the mass and redshift range of  our analysis, we  find that the abundance ratios of QGs do not show any substantial dependence on either mass, size or redshift. This suggests rather universal  formation timescales for the stellar populations of massive QGs of all sizes over a broad range of stellar masses. This implies short gas consumption timescales for the formation of the bulk of the stellar populations. (Note that use of $\alpha$/Fe to infer quenching timescales depends strongly on the amount of mass produced during quenching). The result of \citet{thomas2005}, for a dependence of the abundance ratios on stellar mass, could  nevertheless be recovered considering two points, $(i)$ their analysis extends  down to significantly lower masses, and, also, $(ii)$ that they analyse the $z=0$ population, which will have added a non-negligible number of galaxies at the masses of our study in the intervening several  billion years.

The constancy of the [$\alpha$/Fe] ratios of the QG populations above and below the $10^{11} \mathrm{M}_\odot$ threshold at the redshifts of our study is in contrast to the different behaviors seen for  their age-size relation, and is well explained if the most massive population originates from dry mergers of the less massive population. This further strengthens the argument for a major role of such mergers in leading to the emergence with cosmic time of the most massive QGs in the Universe at masses $>10^{11}\mathrm{M}_\odot$

\section{Summary and Concluding Remarks}\label{conclusion}

The observed average size of QGs is about $\sim3-5$ times larger today than at $z\sim2$. There are two main scenarios which have been proposed to explain this evolution: the size growth of individual galaxies or the progenitor bias introduced if newly formed members of the population are larger than the previous members. In this work, we measure the stellar ages of QGs in order to distinguish between these two scenarios which make quite different predictions for the variation of stellar population age with size.  If the driver of this evolution is the addition of large newly-quenched objects then larger QG galaxies will be younger.  If the driver is the size growth of individual galaxies, and small galaxies are being replaced by newly quenched objects, then the larger galaxies will be older.

 In the light of the fact that, at any epoch and stellar mass,  star-forming galaxies are on average larger than passive galaxies, purity (and completeness) of the QG samples  is clearly crucial when attempting the measurement of  their stellar ages  (see e.g., \citealt{keating2015}).  The polluting presence of star-forming galaxies in QG samples would bias the latter towards larger sizes, while also biasing their age estimates towards younger values. In this work we have selected our QG samples using galaxies securely identified as quiescent. Starting from the 20k zCOSMOS-bright catalog, we selected galaxies with absent, or very weak, emission lines. We stacked the spectra in bins defined in size, stellar mass and redshift, in order to study the average stellar population properties of the sample galaxies.  Two binning schemes were used in size, a relative one normalised to the evolving mean mass-size relation, and an absolute one constructed at fixed physical size. We then used \texttt{pPXF} to derive best-fit ages from BC03 solar SSP templates. To further check our age results, we also computed absorption line strengths following the wavelength definitions of feature and pseudo-continua of the Lick system of spectral line and derived values for age, [Z/H] and [$\alpha$/Fe] comparing our results to the \cite{thomas2011} models. Our robust spectroscopic selection for quenched galaxies is much cleaner than the more frequently adopted color-color selections;
in addition, we carefully checked the spectrum of each object by visual inspection, to ensure the absence of star-formation tracers. We are thus confident that  contamination by  star-forming galaxies is negligible in the results that we have discussed above.

Reassuringly, the average age of QGs (not discriminating in size) increases with cosmic time. Turning to the ages as a function of sizes, we find that the $10^{11} \mathrm{M}_\odot$ mass scale is a `threshold' above and below which the size-age relation changes behavior, as already pointed out in C13. Below $10^{11}{\mathrm{M}}_\odot$,  larger galaxies  have systematically younger ages than smaller ones.  The $\Delta$age between small and large galaxies becomes more significant towards lower redshift. The $\Delta$age from the highest to lowest redshift bin in the small-size QG population  is in  good agreement with a passive evolution of its stellar populations. The younger ages of the larger galaxies at each redshift argues for  newly quenched objects to be  systematically larger at later epochs. This trend is visible using both our size binning schemes.  We conclude that progenitor bias is a major and possibly the dominant component of the observed evolution in the average sizes of QG at these masses. 

Above $10^{11}{\mathrm{M}}_\odot$, where dry mergers are expected to play a major role in imprinting the well-known `dissipationless' features that are observed at $z=0$ in this ultra-massive population, there is indeed no clear trend between ages and sizes.  Size growth of individual galaxies through dry mergers is the most likely channel for the observed growth of the average size of the QG population at this top-mass end.

The confirmation of a `transition' mass around $10^{11}{\mathrm{M}}_\odot$ for  the size-age behaviour -- and thus for the dominant role of progenitor bias at low masses and dry mergers at high masses  in driving  the observed average size growth of QGs with time -- highlights the fundamental importance of sample selection and of tuning the interpretation of the data to the specific sample selection. For example,  \citet{zanella2016} report that the size-age relation in their sample of, in quotation, $\mathrm{M} >4.5\times10^{10}{\mathrm{M}}_\odot$ QGs at $z \sim 1.5$ supports the mergers interpretation. A quick inspection of their analysis shows however that $\sim80\%$ of their galaxies actually have masses above $10^{11} \mathrm{M}_\odot$. Therefore, their result is better commented on as holding for this top-mass end population, which puts their result in agreement with our work. 

Interestingly, the $\alpha$-to-iron abundance ratio of the stellar populations of QGs at all masses within the $10^{10.5-11.5} \mathrm{M}_\odot$ window is rather constant since $z=0.6$. This ratio  should  reflect the formation  timescales for the stellar populations in these systems. The constancy of the measured $[\alpha/\mathrm{Fe}]$ ratio thus suggests similar  such timescales, independent of galaxy size, across the whole  $10^{10.5-11.5} \mathrm{M}_\odot$ mass range for the galaxy population that has already quenched by our lowest redshift bin at $z \sim 0.3$, consistent with the idea that the most massive galaxies above $10^{11} \mathrm{M}_\odot$ are formed by mergers of lower mass galaxies.

\acknowledgments

We acknowledge support from the Swiss National Science Foundation.

\bibliographystyle{apj}
\bibliography{master}

\begin{thebibliography}{}
\expandafter\ifx\csname natexlab\endcsname\relax\def\natexlab#1{#1}\fi

\bibitem[{{Balogh} {et~al.}(1999){Balogh}, {Morris}, {Yee}, {Carlberg}, \&
  {Ellingson}}]{balogh1999}
{Balogh}, M.~L., {Morris}, S.~L., {Yee}, H.~K.~C., {Carlberg}, R.~G., \&
  {Ellingson}, E. 1999, \apj, 527, 54

\bibitem[{{Beasley} {et~al.}(2015){Beasley}, {San Roman}, {Gallart},
  {Sarajedini}, \& {Aparicio}}]{beasley2015}
{Beasley}, M.~A., {San Roman}, I., {Gallart}, C., {Sarajedini}, A., \&
  {Aparicio}, A. 2015, \mnras, 451, 3400

\bibitem[{{Belli} {et~al.}(2015){Belli}, {Newman}, \& {Ellis}}]{belli2015}
{Belli}, S., {Newman}, A.~B., \& {Ellis}, R.~S. 2015, \apj, 799, 206

\bibitem[{{Belli} {et~al.}(2014){Belli}, {Newman}, {Ellis}, \&
  {Konidaris}}]{belli2014b}
{Belli}, S., {Newman}, A.~B., {Ellis}, R.~S., \& {Konidaris}, N.~P. 2014,
  \apjl, 788, L29

\bibitem[{{Bender}(1988)}]{bender1988}
{Bender}, R. 1988, \aap, 202, L5

\bibitem[{{Bertin} \& {Arnouts}(1996)}]{bertin1996}
{Bertin}, E., \& {Arnouts}, S. 1996, \aaps, 117, 393

\bibitem[{{Bruzual} \& {Charlot}(2003)}]{bruzual2003}
{Bruzual}, G., \& {Charlot}, S. 2003, \mnras, 344, 1000

\bibitem[{{Buitrago} {et~al.}(2008){Buitrago}, {Trujillo}, {Conselice},
  {Bouwens}, {Dickinson}, \& {Yan}}]{buitrago2008}
{Buitrago}, F., {Trujillo}, I., {Conselice}, C.~J., {et~al.} 2008, \apjl, 687,
  L61

\bibitem[{{Burstein} {et~al.}(1984){Burstein}, {Faber}, {Gaskell}, \&
  {Krumm}}]{burstein1984}
{Burstein}, D., {Faber}, S.~M., {Gaskell}, C.~M., \& {Krumm}, N. 1984, \apj,
  287, 586

\bibitem[{{Calzetti} {et~al.}(2000){Calzetti}, {Armus}, {Bohlin}, {Kinney},
  {Koornneef}, \& {Storchi-Bergmann}}]{calzetti2000}
{Calzetti}, D., {Armus}, L., {Bohlin}, R.~C., {et~al.} 2000, apj, 533, 682

\bibitem[{{Cappellari}(2013)}]{cappellari2013a}
{Cappellari}, M. 2013, \apjl, 778, L2

\bibitem[{{Cappellari} \& {Emsellem}(2004)}]{cappellari2004}
{Cappellari}, M., \& {Emsellem}, E. 2004, \pasp, 116, 138

\bibitem[{{Cappellari} {et~al.}(2009){Cappellari}, {di Serego Alighieri},
  {Cimatti}, {Daddi}, {Renzini}, {Kurk}, {Cassata}, {Dickinson},
  {Franceschini}, {Mignoli}, {Pozzetti}, {Rodighiero}, {Rosati}, \&
  {Zamorani}}]{cappellari2009}
{Cappellari}, M., {di Serego Alighieri}, S., {Cimatti}, A., {et~al.} 2009,
  \apjl, 704, L34

\bibitem[{{Caputi} {et~al.}(2008){Caputi}, {Lilly}, {Aussel}, {Sanders},
  {Frayer}, {Le F{\`e}vre}, {Renzini}, {Zamorani}, {Scodeggio}, {Contini},
  {Scoville}, {Carollo}, {Hasinger}, {Iovino}, {Le Brun}, {Le Floc'h}, {Maier},
  {Mainieri}, {Mignoli}, {Salvato}, {Schiminovich}, {Silverman}, {Surace},
  {Tasca}, {Abbas}, {Bardelli}, {Bolzonella}, {Bongiorno}, {Bottini}, {Capak},
  {Cappi}, {Cassata}, {Cimatti}, {Cucciati}, {de la Torre}, {de Ravel},
  {Franzetti}, {Fumana}, {Garilli}, {Halliday}, {Ilbert}, {Kampczyk},
  {Kartaltepe}, {Kneib}, {Knobel}, {Kovac}, {Lamareille}, {Leauthaud}, {Le
  Borgne}, {Maccagni}, {Marinoni}, {McCracken}, {Meneux}, {Oesch}, {Pell{\`o}},
  {P{\'e}rez-Montero}, {Porciani}, {Ricciardelli}, {Scaramella}, {Scarlata},
  {Tresse}, {Vergani}, {Walcher}, {Zamojski}, \& {Zucca}}]{caputi2008}
{Caputi}, K.~I., {Lilly}, S.~J., {Aussel}, H., {et~al.} 2008, \apj, 680, 939

\bibitem[{{Carollo} \& {Danziger}(1994)}]{carollo1994}
{Carollo}, C.~M., \& {Danziger}, I.~J. 1994, \mnras, 270, 523

\bibitem[{{Carollo} {et~al.}(1993){Carollo}, {Danziger}, \&
  {Buson}}]{carollo1993}
{Carollo}, C.~M., {Danziger}, I.~J., \& {Buson}, L. 1993, \mnras, 265, 553

\bibitem[{{Carollo} {et~al.}(2013){Carollo}, {Bschorr}, {Renzini}, {Lilly},
  {Capak}, {Cibinel}, {Ilbert}, {Onodera}, {Scoville}, {Cameron}, {Mobasher},
  {Sanders}, \& {Taniguchi}}]{carollo2013}
{Carollo}, C.~M., {Bschorr}, T.~J., {Renzini}, A., {et~al.} 2013, \apj, 773,
  112

\bibitem[{{Carollo} {et~al.}(2016){Carollo}, {Cibinel}, {Lilly}, {Pipino},
  {Bonoli}, {Finoguenov}, {Miniati}, {Norberg}, \& {Silverman}}]{carollo2016}
{Carollo}, C.~M., {Cibinel}, A., {Lilly}, S.~J., {et~al.} 2016, \apj, 818, 180

\bibitem[{{Cassata} {et~al.}(2011){Cassata}, {Giavalisco}, {Guo}, {Renzini},
  {Ferguson}, {Koekemoer}, {Salimbeni}, {Scarlata}, {Grogin}, {Conselice},
  {Dahlen}, {Lotz}, {Dickinson}, \& {Lin}}]{cassata2011}
{Cassata}, P., {Giavalisco}, M., {Guo}, Y., {et~al.} 2011, \apj, 743, 96

\bibitem[{{Cassata} {et~al.}(2013){Cassata}, {Giavalisco}, {Williams}, {Guo},
  {Lee}, {Renzini}, {Ferguson}, {Faber}, {Barro}, {McIntosh}, {Lu}, {Bell},
  {Koo}, {Papovich}, {Ryan}, {Conselice}, {Grogin}, {Koekemoer}, \&
  {Hathi}}]{cassata2013}
{Cassata}, P., {Giavalisco}, M., {Williams}, C.~C., {et~al.} 2013, \apj, 775,
  106

\bibitem[{{Chabrier}(2003)}]{chabrier2003}
{Chabrier}, G. 2003, \pasp, 115, 763

\bibitem[{{Choi} {et~al.}(2014){Choi}, {Conroy}, {Moustakas}, {Graves},
  {Holden}, {Brodwin}, {Brown}, \& {van Dokkum}}]{choi2014}
{Choi}, J., {Conroy}, C., {Moustakas}, J., {et~al.} 2014, \apj, 792, 95

\bibitem[{{Cimatti} {et~al.}(2008){Cimatti}, {Cassata}, {Pozzetti}, {Kurk},
  {Mignoli}, {Renzini}, {Daddi}, {Bolzonella}, {Brusa}, {Rodighiero},
  {Dickinson}, {Franceschini}, {Zamorani}, {Berta}, {Rosati}, \&
  {Halliday}}]{cimatti2008}
{Cimatti}, A., {Cassata}, P., {Pozzetti}, L., {et~al.} 2008, \aap, 482, 21

\bibitem[{{Conroy} \& {van Dokkum}(2012)}]{conroy2012}
{Conroy}, C., \& {van Dokkum}, P.~G. 2012, \apj, 760, 71

\bibitem[{{Daddi} {et~al.}(2005){Daddi}, {Renzini}, {Pirzkal}, {Cimatti},
  {Malhotra}, {Stiavelli}, {Xu}, {Pasquali}, {Rhoads}, {Brusa}, {di Serego
  Alighieri}, {Ferguson}, {Koekemoer}, {Moustakas}, {Panagia}, \&
  {Windhorst}}]{daddi2005}
{Daddi}, E., {Renzini}, A., {Pirzkal}, N., {et~al.} 2005, \apj, 626, 680

\bibitem[{{Damjanov} {et~al.}(2015){Damjanov}, {Geller}, {Zahid}, \&
  {Hwang}}]{damjanov2015a}
{Damjanov}, I., {Geller}, M.~J., {Zahid}, H.~J., \& {Hwang}, H.~S. 2015, \apj,
  806, 158

\bibitem[{{Dom{\'{\i}}nguez S{\'a}nchez} {et~al.}(2011){Dom{\'{\i}}nguez
  S{\'a}nchez}, {Pozzi}, {Gruppioni}, {Cimatti}, {Ilbert}, {Pozzetti},
  {McCracken}, {Capak}, {Le Floch}, {Salvato}, {Zamorani}, {Carollo},
  {Contini}, {Kneib}, {Le F{\`e}vre}, {Lilly}, {Mainieri}, {Renzini},
  {Scodeggio}, {Bardelli}, {Bolzonella}, {Bongiorno}, {Caputi}, {Coppa},
  {Cucciati}, {de la Torre}, {de Ravel}, {Franzetti}, {Garilli}, {Iovino},
  {Kampczyk}, {Knobel}, {Kova{\v c}}, {Lamareille}, {Le Borgne}, {Le Brun},
  {Maier}, {Mignoli}, {Pell{\'o}}, {Peng}, {Perez-Montero}, {Ricciardelli},
  {Silverman}, {Tanaka}, {Tasca}, {Tresse}, {Vergani}, \&
  {Zucca}}]{dominguezsanchez2011}
{Dom{\'{\i}}nguez S{\'a}nchez}, H., {Pozzi}, F., {Gruppioni}, C., {et~al.}
  2011, \mnras, 417, 900

\bibitem[{{Elvis} {et~al.}(2009){Elvis}, {Civano}, {Vignali}, {Puccetti},
  {Fiore}, {Cappelluti}, {Aldcroft}, {Fruscione}, {Zamorani}, {Comastri},
  {Brusa}, {Gilli}, {Miyaji}, {Damiani}, {Koekemoer}, {Finoguenov}, {Brunner},
  {Urry}, {Silverman}, {Mainieri}, {Hasinger}, {Griffiths}, {Carollo}, {Hao},
  {Guzzo}, {Blain}, {Calzetti}, {Carilli}, {Capak}, {Ettori}, {Fabbiano},
  {Impey}, {Lilly}, {Mobasher}, {Rich}, {Salvato}, {Sanders}, {Schinnerer},
  {Scoville}, {Shopbell}, {Taylor}, {Taniguchi}, \& {Volonteri}}]{elvis2009}
{Elvis}, M., {Civano}, F., {Vignali}, C., {et~al.} 2009, \apjs, 184, 158

\bibitem[{{Faber} {et~al.}(2007){Faber}, {Willmer}, {Wolf}, {Koo}, {Weiner},
  {Newman}, {Im}, {Coil}, {Conroy}, {Cooper}, {Davis}, {Finkbeiner}, {Gerke},
  {Gebhardt}, {Groth}, {Guhathakurta}, {Harker}, {Kaiser}, {Kassin},
  {Kleinheinrich}, {Konidaris}, {Kron}, {Lin}, {Luppino}, {Madgwick},
  {Meisenheimer}, {Noeske}, {Phillips}, {Sarajedini}, {Schiavon}, {Simard},
  {Szalay}, {Vogt}, \& {Yan}}]{faber2007}
{Faber}, S.~M., {Willmer}, C.~N.~A., {Wolf}, C., {et~al.} 2007, \apj, 665, 265

\bibitem[{{Falc{\'o}n-Barroso} {et~al.}(2011){Falc{\'o}n-Barroso},
  {S{\'a}nchez-Bl{\'a}zquez}, {Vazdekis}, {Ricciardelli}, {Cardiel}, {Cenarro},
  {Gorgas}, \& {Peletier}}]{falconbarroso2011}
{Falc{\'o}n-Barroso}, J., {S{\'a}nchez-Bl{\'a}zquez}, P., {Vazdekis}, A.,
  {et~al.} 2011, aap, 532, A95

\bibitem[{{Feldmann} {et~al.}(2010){Feldmann}, {Carollo}, {Mayer}, {Renzini},
  {Lake}, {Quinn}, {Stinson}, \& {Yepes}}]{feldmann2010}
{Feldmann}, R., {Carollo}, C.~M., {Mayer}, L., {et~al.} 2010, \apj, 709, 218

\bibitem[{{Fisher} {et~al.}(1996){Fisher}, {Franx}, \&
  {Illingworth}}]{fisher1996}
{Fisher}, D., {Franx}, M., \& {Illingworth}, G. 1996, \apj, 459, 110

\bibitem[{{Franx} {et~al.}(2008){Franx}, {van Dokkum}, {Schreiber}, {Wuyts},
  {Labb{\'e}}, \& {Toft}}]{franx2008}
{Franx}, M., {van Dokkum}, P.~G., {Schreiber}, N.~M.~F., {et~al.} 2008, \apj,
  688, 770

\bibitem[{{Gallazzi} {et~al.}(2014){Gallazzi}, {Bell}, {Zibetti}, {Brinchmann},
  \& {Kelson}}]{gallazzi2014}
{Gallazzi}, A., {Bell}, E.~F., {Zibetti}, S., {Brinchmann}, J., \& {Kelson},
  D.~D. 2014, \apj, 788, 72

\bibitem[{{Gallazzi} {et~al.}(2005){Gallazzi}, {Charlot}, {Brinchmann},
  {White}, \& {Tremonti}}]{gallazzi2005}
{Gallazzi}, A., {Charlot}, S., {Brinchmann}, J., {White}, S.~D.~M., \&
  {Tremonti}, C.~A. 2005, \mnras, 362, 41

\bibitem[{{Genzel} {et~al.}(2015){Genzel}, {Tacconi}, {Lutz}, {Saintonge},
  {Berta}, {Magnelli}, {Combes}, {Garc{\'{\i}}a-Burillo}, {Neri}, {Bolatto},
  {Contini}, {Lilly}, {Boissier}, {Boone}, {Bouch{\'e}}, {Bournaud}, {Burkert},
  {Carollo}, {Colina}, {Cooper}, {Cox}, {Feruglio}, {F{\"o}rster Schreiber},
  {Freundlich}, {Gracia-Carpio}, {Juneau}, {Kovac}, {Lippa}, {Naab}, {Salome},
  {Renzini}, {Sternberg}, {Walter}, {Weiner}, {Weiss}, \& {Wuyts}}]{genzel2015}
{Genzel}, R., {Tacconi}, L.~J., {Lutz}, D., {et~al.} 2015, \apj, 800, 20

\bibitem[{{Gonz{\'a}lez}(1993)}]{gonzalez1993}
{Gonz{\'a}lez}, J.~J. 1993, PhD thesis, Thesis (PH.D.)--UNIVERSITY OF
  CALIFORNIA, SANTA CRUZ, 1993.Source: Dissertation Abstracts International,
  Volume: 54-05, Section: B, page: 2551.

\bibitem[{{Guzzo} {et~al.}(2007){Guzzo}, {Cassata}, {Finoguenov}, {Massey},
  {Scoville}, {Capak}, {Ellis}, {Mobasher}, {Taniguchi}, {Thompson}, {Ajiki},
  {Aussel}, {B{\"o}hringer}, {Brusa}, {Calzetti}, {Comastri}, {Franceschini},
  {Hasinger}, {Kasliwal}, {Kitzbichler}, {Kneib}, {Koekemoer}, {Leauthaud},
  {McCracken}, {Murayama}, {Nagao}, {Rhodes}, {Sanders}, {Sasaki}, {Shioya},
  {Tasca}, \& {Taylor}}]{guzzo2007}
{Guzzo}, L., {Cassata}, P., {Finoguenov}, A., {et~al.} 2007, \apjs, 172, 254

\bibitem[{{Hilz} {et~al.}(2012){Hilz}, {Naab}, {Ostriker}, {Thomas}, {Burkert},
  \& {Jesseit}}]{hilz2012}
{Hilz}, M., {Naab}, T., {Ostriker}, J.~P., {et~al.} 2012, \mnras, 425, 3119

\bibitem[{{Hopkins} {et~al.}(2009){Hopkins}, {Bundy}, {Murray}, {Quataert},
  {Lauer}, \& {Ma}}]{hopkins2009}
{Hopkins}, P.~F., {Bundy}, K., {Murray}, N., {et~al.} 2009, \mnras, 398, 898

\bibitem[{{Ilbert} {et~al.}(2010){Ilbert}, {Salvato}, {Le Floc'h}, {Aussel},
  {Capak}, {McCracken}, {Mobasher}, {Kartaltepe}, {Scoville}, {Sanders},
  {Arnouts}, {Bundy}, {Cassata}, {Kneib}, {Koekemoer}, {Le F{\`e}vre}, {Lilly},
  {Surace}, {Taniguchi}, {Tasca}, {Thompson}, {Tresse}, {Zamojski}, {Zamorani},
  \& {Zucca}}]{ilbert2010}
{Ilbert}, O., {Salvato}, M., {Le Floc'h}, E., {et~al.} 2010, \apj, 709, 644

\bibitem[{{Ilbert} {et~al.}(2013){Ilbert}, {McCracken}, {Le F{\`e}vre},
  {Capak}, {Dunlop}, {Karim}, {Renzini}, {Caputi}, {Boissier}, {Arnouts},
  {Aussel}, {Comparat}, {Guo}, {Hudelot}, {Kartaltepe}, {Kneib}, {Krogager},
  {Le Floc'h}, {Lilly}, {Mellier}, {Milvang-Jensen}, {Moutard}, {Onodera},
  {Richard}, {Salvato}, {Sanders}, {Scoville}, {Silverman}, {Taniguchi},
  {Tasca}, {Thomas}, {Toft}, {Tresse}, {Vergani}, {Wolk}, \&
  {Zirm}}]{ilbert2013}
{Ilbert}, O., {McCracken}, H.~J., {Le F{\`e}vre}, O., {et~al.} 2013, \aap, 556,
  A55

\bibitem[{{Keating} {et~al.}(2015){Keating}, {Abraham}, {Schiavon}, {Graves},
  {Damjanov}, {Yan}, {Newman}, \& {Simard}}]{keating2015}
{Keating}, S.~K., {Abraham}, R.~G., {Schiavon}, R., {et~al.} 2015, \apj, 798,
  26

\bibitem[{{Kelson} {et~al.}(2000){Kelson}, {Illingworth}, {van Dokkum}, \&
  {Franx}}]{Kelson2000}
{Kelson}, D.~D., {Illingworth}, G.~D., {van Dokkum}, P.~G., \& {Franx}, M.
  2000, \apj, 531, 159

\bibitem[{{Knobel} {et~al.}(2009){Knobel}, {Lilly}, {Iovino}, {Porciani},
  {Kova{\v c}}, {Cucciati}, {Finoguenov}, {Kitzbichler}, {Carollo}, {Contini},
  {Kneib}, {Le F{\`e}vre}, {Mainieri}, {Renzini}, {Scodeggio}, {Zamorani},
  {Bardelli}, {Bolzonella}, {Bongiorno}, {Caputi}, {Coppa}, {de la Torre}, {de
  Ravel}, {Franzetti}, {Garilli}, {Kampczyk}, {Lamareille}, {Le Borgne}, {Le
  Brun}, {Maier}, {Mignoli}, {Pello}, {Peng}, {Perez Montero}, {Ricciardelli},
  {Silverman}, {Tanaka}, {Tasca}, {Tresse}, {Vergani}, {Zucca}, {Abbas},
  {Bottini}, {Cappi}, {Cassata}, {Cimatti}, {Fumana}, {Guzzo}, {Koekemoer},
  {Leauthaud}, {Maccagni}, {Marinoni}, {McCracken}, {Memeo}, {Meneux}, {Oesch},
  {Pozzetti}, \& {Scaramella}}]{knobel2009}
{Knobel}, C., {Lilly}, S.~J., {Iovino}, A., {et~al.} 2009, \apj, 697, 1842

\bibitem[{{Koekemoer} {et~al.}(2007){Koekemoer}, {Aussel}, {Calzetti}, {Capak},
  {Giavalisco}, {Kneib}, {Leauthaud}, {Le F{\`e}vre}, {McCracken}, {Massey},
  {Mobasher}, {Rhodes}, {Scoville}, \& {Shopbell}}]{Koekemoer2007}
{Koekemoer}, A.~M., {Aussel}, H., {Calzetti}, D., {et~al.} 2007, \apjs, 172,
  196

\bibitem[{{Koleva} {et~al.}(2009){Koleva}, {Prugniel}, {Bouchard}, \&
  {Wu}}]{koleva2009}
{Koleva}, M., {Prugniel}, P., {Bouchard}, A., \& {Wu}, Y. 2009, \aap, 501, 1269

\bibitem[{{Koleva} {et~al.}(2008){Koleva}, {Prugniel}, {Ocvirk}, {Le Borgne},
  \& {Soubiran}}]{koleva2008}
{Koleva}, M., {Prugniel}, P., {Ocvirk}, P., {Le Borgne}, D., \& {Soubiran}, C.
  2008, \mnras, 385, 1998

\bibitem[{{Korn} {et~al.}(2005){Korn}, {Maraston}, \& {Thomas}}]{korn2005}
{Korn}, A.~J., {Maraston}, C., \& {Thomas}, D. 2005, \aap, 438, 685

\bibitem[{{Kriek} {et~al.}(2009){Kriek}, {van Dokkum}, {Labb{\'e}}, {Franx},
  {Illingworth}, {Marchesini}, \& {Quadri}}]{kriek2009}
{Kriek}, M., {van Dokkum}, P.~G., {Labb{\'e}}, I., {et~al.} 2009, \apj, 700,
  221

\bibitem[{{Kron}(1980)}]{kron1980}
{Kron}, R.~G. 1980, \apjs, 43, 305

\bibitem[{{Kuntschner} {et~al.}(2001){Kuntschner}, {Lucey}, {Smith}, {Hudson},
  \& {Davies}}]{kuntschner2001}
{Kuntschner}, H., {Lucey}, J.~R., {Smith}, R.~J., {Hudson}, M.~J., \& {Davies},
  R.~L. 2001, \mnras, 323, 615

\bibitem[{{Lee} \& {Worthey}(2005)}]{leeworthey2005}
{Lee}, H.-c., \& {Worthey}, G. 2005, \apjs, 160, 176

\bibitem[{{Lilly} \& {Carollo}(2016)}]{lilly2016}
{Lilly}, S.~J., \& {Carollo}, C.~M. 2016, ArXiv e-prints, arXiv:1604.06459

\bibitem[{{Lilly} {et~al.}(2007){Lilly}, {Le F{\`e}vre}, {Renzini}, {Zamorani},
  {Scodeggio}, {Contini}, {Carollo}, {Hasinger}, {Kneib}, {Iovino}, {Le Brun},
  {Maier}, {Mainieri}, {Mignoli}, {Silverman}, {Tasca}, {Bolzonella},
  {Bongiorno}, {Bottini}, {Capak}, {Caputi}, {Cimatti}, {Cucciati}, {Daddi},
  {Feldmann}, {Franzetti}, {Garilli}, {Guzzo}, {Ilbert}, {Kampczyk}, {Kovac},
  {Lamareille}, {Leauthaud}, {Borgne}, {McCracken}, {Marinoni}, {Pello},
  {Ricciardelli}, {Scarlata}, {Vergani}, {Sanders}, {Schinnerer}, {Scoville},
  {Taniguchi}, {Arnouts}, {Aussel}, {Bardelli}, {Brusa}, {Cappi}, {Ciliegi},
  {Finoguenov}, {Foucaud}, {Franceschini}, {Halliday}, {Impey}, {Knobel},
  {Koekemoer}, {Kurk}, {Maccagni}, {Maddox}, {Marano}, {Marconi}, {Meneux},
  {Mobasher}, {Moreau}, {Peacock}, {Porciani}, {Pozzetti}, {Scaramella},
  {Schiminovich}, {Shopbell}, {Smail}, {Thompson}, {Tresse}, {Vettolani},
  {Zanichelli}, \& {Zucca}}]{lilly2007}
{Lilly}, S.~J., {Le F{\`e}vre}, O., {Renzini}, A., {et~al.} 2007, \apjs, 172,
  70

\bibitem[{{Lilly} {et~al.}(2009){Lilly}, {Le Brun}, {Maier}, {Mainieri},
  {Mignoli}, {Scodeggio}, {Zamorani}, {Carollo}, {Contini}, {Kneib}, {Le
  F{\`e}vre}, {Renzini}, {Bardelli}, {Bolzonella}, {Bongiorno}, {Caputi},
  {Coppa}, {Cucciati}, {de la Torre}, {de Ravel}, {Franzetti}, {Garilli},
  {Iovino}, {Kampczyk}, {Kovac}, {Knobel}, {Lamareille}, {Le Borgne}, {Pello},
  {Peng}, {P{\'e}rez-Montero}, {Ricciardelli}, {Silverman}, {Tanaka}, {Tasca},
  {Tresse}, {Vergani}, {Zucca}, {Ilbert}, {Salvato}, {Oesch}, {Abbas},
  {Bottini}, {Capak}, {Cappi}, {Cassata}, {Cimatti}, {Elvis}, {Fumana},
  {Guzzo}, {Hasinger}, {Koekemoer}, {Leauthaud}, {Maccagni}, {Marinoni},
  {McCracken}, {Memeo}, {Meneux}, {Porciani}, {Pozzetti}, {Sanders},
  {Scaramella}, {Scarlata}, {Scoville}, {Shopbell}, \& {Taniguchi}}]{lilly2009}
{Lilly}, S.~J., {Le Brun}, V., {Maier}, C., {et~al.} 2009, \apjs, 184, 218

\bibitem[{{MacArthur}(2005)}]{macarthur2005}
{MacArthur}, L.~A. 2005, \apj, 623, 795

\bibitem[{{Matteucci} \& {Greggio}(1986)}]{matteucci1986}
{Matteucci}, F., \& {Greggio}, L. 1986, \aap, 154, 279

\bibitem[{{McLure} {et~al.}(2013){McLure}, {Pearce}, {Dunlop}, {Cirasuolo},
  {Curtis-Lake}, {Bruce}, {Caputi}, {Almaini}, {Bonfield}, {Bradshaw},
  {Buitrago}, {Chuter}, {Foucaud}, {Hartley}, \& {Jarvis}}]{mclure2013}
{McLure}, R.~J., {Pearce}, H.~J., {Dunlop}, J.~S., {et~al.} 2013, \mnras, 428,
  1088

\bibitem[{{Mignoli} {et~al.}(2009){Mignoli}, {Zamorani}, {Scodeggio},
  {Cimatti}, {Halliday}, {Lilly}, {Pozzetti}, {Vergani}, {Carollo}, {Contini},
  {Le F{\'e}vre}, {Mainieri}, {Renzini}, {Bardelli}, {Bolzonella}, {Bongiorno},
  {Caputi}, {Coppa}, {Cucciati}, {de La Torre}, {de Ravel}, {Franzetti},
  {Garilli}, {Iovino}, {Kampczyk}, {Kneib}, {Knobel}, {Kova{\v c}},
  {Lamareille}, {Le Borgne}, {Le Brun}, {Maier}, {Pell{\`o}}, {Peng}, {Perez
  Montero}, {Ricciardelli}, {Scarlata}, {Silverman}, {Tanaka}, {Tasca},
  {Tresse}, {Zucca}, {Abbas}, {Bottini}, {Capak}, {Cappi}, {Cassata}, {Fumana},
  {Guzzo}, {Leauthaud}, {Maccagni}, {Marinoni}, {McCracken}, {Memeo}, {Meneux},
  {Oesch}, {Porciani}, {Scaramella}, \& {Scoville}}]{mignoli2009}
{Mignoli}, M., {Zamorani}, G., {Scodeggio}, M., {et~al.} 2009, \aap, 493, 39

\bibitem[{{Moresco} {et~al.}(2010){Moresco}, {Pozzetti}, {Cimatti}, {Zamorani},
  {Mignoli}, {di Cesare}, {Bolzonella}, {Zucca}, {Lilly}, {Kova{\v c}},
  {Scodeggio}, {Cassata}, {Tasca}, {Vergani}, {Halliday}, {Carollo}, {Contini},
  {Kneib}, {Le F{\'e}vre}, {Mainieri}, {Renzini}, {Bardelli}, {Bongiorno},
  {Caputi}, {Coppa}, {Cucciati}, {de la Torre}, {de Ravel}, {Franzetti},
  {Garilli}, {Iovino}, {Kampczyk}, {Knobel}, {Lamareille}, {Le Borgne}, {Le
  Brun}, {Maier}, {Pell{\`o}}, {Peng}, {Perez Montero}, {Ricciardelli},
  {Silverman}, {Tanaka}, {Tresse}, {Abbas}, {Bottini}, {Cappi}, {Guzzo},
  {Koekemoer}, {Leauthaud}, {Maccagni}, {Marinoni}, {McCracken}, {Memeo},
  {Meneux}, {Nair}, {Oesch}, {Porciani}, {Scaramella}, {Scarlata}, \&
  {Scoville}}]{moresco2010}
{Moresco}, M., {Pozzetti}, L., {Cimatti}, A., {et~al.} 2010, \aap, 524, A67

\bibitem[{{Moresco} {et~al.}(2013){Moresco}, {Pozzetti}, {Cimatti}, {Zamorani},
  {Bolzonella}, {Lamareille}, {Mignoli}, {Zucca}, {Lilly}, {Carollo},
  {Contini}, {Kneib}, {Le F{\`e}vre}, {Mainieri}, {Renzini}, {Scodeggio},
  {Bardelli}, {Bongiorno}, {Caputi}, {Cucciati}, {de la Torre}, {de Ravel},
  {Franzetti}, {Garilli}, {Iovino}, {Kampczyk}, {Knobel}, {Kova{\v c}}, {Le
  Borgne}, {Le Brun}, {Maier}, {Pell{\'o}}, {Peng}, {Perez-Montero},
  {Presotto}, {Silverman}, {Tanaka}, {Tasca}, {Tresse}, {Vergani}, {Barnes},
  {Bordoloi}, {Cappi}, {Diener}, {Koekemoer}, {Le Floc'h}, {L{\'o}pez-Sanjuan},
  {McCracken}, {Nair}, {Oesch}, {Scarlata}, {Scoville}, \&
  {Welikala}}]{moresco2013}
---. 2013, \aap, 558, A61

\bibitem[{{Mosleh} {et~al.}(2013){Mosleh}, {Williams}, \& {Franx}}]{mosleh2013}
{Mosleh}, M., {Williams}, R.~J., \& {Franx}, M. 2013, \apj, 777, 117

\bibitem[{{Muzzin} {et~al.}(2013){Muzzin}, {Marchesini}, {Stefanon}, {Franx},
  {Milvang-Jensen}, {Dunlop}, {Fynbo}, {Brammer}, {Labb{\'e}}, \& {van
  Dokkum}}]{muzzin2013}
{Muzzin}, A., {Marchesini}, D., {Stefanon}, M., {et~al.} 2013, /apjs, 206, 8

\bibitem[{{Naab} {et~al.}(2009){Naab}, {Johansson}, \& {Ostriker}}]{naab2009}
{Naab}, T., {Johansson}, P.~H., \& {Ostriker}, J.~P. 2009, \apjl, 699, L178

\bibitem[{{Newman} {et~al.}(2012){Newman}, {Ellis}, {Bundy}, \&
  {Treu}}]{newman2012}
{Newman}, A.~B., {Ellis}, R.~S., {Bundy}, K., \& {Treu}, T. 2012, \apj, 746,
  162

\bibitem[{{Nomoto} {et~al.}(1984){Nomoto}, {Thielemann}, \&
  {Wheeler}}]{nomoto1984}
{Nomoto}, K., {Thielemann}, F.-K., \& {Wheeler}, J.~C. 1984, \apjl, 279, L23

\bibitem[{{Ocvirk} {et~al.}(2006{\natexlab{a}}){Ocvirk}, {Pichon}, {Lan{\c
  c}on}, \& {Thi{\'e}baut}}]{ocvirk2006b}
{Ocvirk}, P., {Pichon}, C., {Lan{\c c}on}, A., \& {Thi{\'e}baut}, E.
  2006{\natexlab{a}}, \mnras, 365, 74

\bibitem[{{Ocvirk} {et~al.}(2006{\natexlab{b}}){Ocvirk}, {Pichon}, {Lan{\c
  c}on}, \& {Thi{\'e}baut}}]{ocvirk2006a}
---. 2006{\natexlab{b}}, \mnras, 365, 46

\bibitem[{{O'Donnell}(1994)}]{odonnell1994}
{O'Donnell}, J.~E. 1994, \apj, 422, 158

\bibitem[{{Oesch} {et~al.}(2010){Oesch}, {Carollo}, {Feldmann}, {Hahn},
  {Lilly}, {Sargent}, {Scarlata}, {Aller}, {Aussel}, {Bolzonella}, {Bschorr},
  {Bundy}, {Capak}, {Ilbert}, {Kneib}, {Koekemoer}, {Kova{\v c}}, {Leauthaud},
  {Le Floc'h}, {Massey}, {McCracken}, {Pozzetti}, {Renzini}, {Rhodes},
  {Salvato}, {Sanders}, {Scoville}, {Sheth}, {Taniguchi}, \&
  {Thompson}}]{oesch2010}
{Oesch}, P.~A., {Carollo}, C.~M., {Feldmann}, R., {et~al.} 2010, \apjl, 714,
  L47

\bibitem[{{Onodera} {et~al.}(2012){Onodera}, {Renzini}, {Carollo},
  {Cappellari}, {Mancini}, {Strazzullo}, {Daddi}, {Arimoto}, {Gobat}, {Yamada},
  {McCracken}, {Ilbert}, {Capak}, {Cimatti}, {Giavalisco}, {Koekemoer}, {Kong},
  {Lilly}, {Motohara}, {Ohta}, {Sanders}, {Scoville}, {Tamura}, \&
  {Taniguchi}}]{onodera2012}
{Onodera}, M., {Renzini}, A., {Carollo}, M., {et~al.} 2012, \apj, 755, 26

\bibitem[{{Onodera} {et~al.}(2015){Onodera}, {Carollo}, {Renzini},
  {Cappellari}, {Mancini}, {Arimoto}, {Daddi}, {Gobat}, {Strazzullo},
  {Tacchella}, \& {Yamada}}]{onodera2014}
{Onodera}, M., {Carollo}, C.~M., {Renzini}, A., {et~al.} 2015, \apj, 808, 161

\bibitem[{Osterbrock \& Ferland(2006)}]{osterbrock2006}
Osterbrock, D., \& Ferland, G. 2006, Astrophysics of Gaseous Nebulae and Active
  Galactic Nuclei (University Science Books)

\bibitem[{{Pagel} \& {Tautvaisiene}(1995)}]{pagel1995}
{Pagel}, B.~E.~J., \& {Tautvaisiene}, G. 1995, \mnras, 276, 505

\bibitem[{{Patel} {et~al.}(2009){Patel}, {Holden}, {Kelson}, {Illingworth}, \&
  {Franx}}]{patel2009}
{Patel}, S.~G., {Holden}, B.~P., {Kelson}, D.~D., {Illingworth}, G.~D., \&
  {Franx}, M. 2009, apjl, 705, L67

\bibitem[{{Patel} {et~al.}(2013){Patel}, {van Dokkum}, {Franx}, {Quadri},
  {Muzzin}, {Marchesini}, {Williams}, {Holden}, \& {Stefanon}}]{patel2013}
{Patel}, S.~G., {van Dokkum}, P.~G., {Franx}, M., {et~al.} 2013, \apj, 766, 15

\bibitem[{{Peng} {et~al.}(2010){Peng}, {Lilly}, {Kova{\v c}}, {Bolzonella},
  {Pozzetti}, {Renzini}, {Zamorani}, {Ilbert}, {Knobel}, {Iovino}, {Maier},
  {Cucciati}, {Tasca}, {Carollo}, {Silverman}, {Kampczyk}, {de Ravel},
  {Sanders}, {Scoville}, {Contini}, {Mainieri}, {Scodeggio}, {Kneib}, {Le
  F{\`e}vre}, {Bardelli}, {Bongiorno}, {Caputi}, {Coppa}, {de la Torre},
  {Franzetti}, {Garilli}, {Lamareille}, {Le Borgne}, {Le Brun}, {Mignoli},
  {Perez Montero}, {Pello}, {Ricciardelli}, {Tanaka}, {Tresse}, {Vergani},
  {Welikala}, {Zucca}, {Oesch}, {Abbas}, {Barnes}, {Bordoloi}, {Bottini},
  {Cappi}, {Cassata}, {Cimatti}, {Fumana}, {Hasinger}, {Koekemoer},
  {Leauthaud}, {Maccagni}, {Marinoni}, {McCracken}, {Memeo}, {Meneux}, {Nair},
  {Porciani}, {Presotto}, \& {Scaramella}}]{peng2010}
{Peng}, Y.-j., {Lilly}, S.~J., {Kova{\v c}}, K., {et~al.} 2010, \apj, 721, 193

\bibitem[{{Poggianti} {et~al.}(2013){Poggianti}, {Moretti}, {Calvi},
  {D'Onofrio}, {Valentinuzzi}, {Fritz}, \& {Renzini}}]{poggianti2013}
{Poggianti}, B.~M., {Moretti}, A., {Calvi}, R., {et~al.} 2013, \apj, 777, 125

\bibitem[{{Poggianti} {et~al.}(2001){Poggianti}, {Bridges}, {Mobasher},
  {Carter}, {Doi}, {Iye}, {Kashikawa}, {Komiyama}, {Okamura}, {Sekiguchi},
  {Shimasaku}, {Yagi}, \& {Yasuda}}]{poggianti2001}
{Poggianti}, B.~M., {Bridges}, T.~J., {Mobasher}, B., {et~al.} 2001, \apj, 562,
  689

\bibitem[{{Poole} {et~al.}(2010){Poole}, {Worthey}, {Lee}, \&
  {Serven}}]{poole2010}
{Poole}, V., {Worthey}, G., {Lee}, H.-c., \& {Serven}, J. 2010, \aj, 139, 809

\bibitem[{{Pozzetti} {et~al.}(2010){Pozzetti}, {Bolzonella}, {Zucca},
  {Zamorani}, {Lilly}, {Renzini}, {Moresco}, {Mignoli}, {Cassata}, {Tasca},
  {Lamareille}, {Maier}, {Meneux}, {Halliday}, {Oesch}, {Vergani}, {Caputi},
  {Kova{\v c}}, {Cimatti}, {Cucciati}, {Iovino}, {Peng}, {Carollo}, {Contini},
  {Kneib}, {Le F{\'e}vre}, {Mainieri}, {Scodeggio}, {Bardelli}, {Bongiorno},
  {Coppa}, {de la Torre}, {de Ravel}, {Franzetti}, {Garilli}, {Kampczyk},
  {Knobel}, {Le Borgne}, {Le Brun}, {Pell{\`o}}, {Perez Montero},
  {Ricciardelli}, {Silverman}, {Tanaka}, {Tresse}, {Abbas}, {Bottini}, {Cappi},
  {Guzzo}, {Koekemoer}, {Leauthaud}, {Maccagni}, {Marinoni}, {McCracken},
  {Memeo}, {Porciani}, {Scaramella}, {Scarlata}, \& {Scoville}}]{pozzetti2010}
{Pozzetti}, L., {Bolzonella}, M., {Zucca}, E., {et~al.} 2010, \aap, 523, A13

\bibitem[{{Renzini}(2006)}]{renzini2006}
{Renzini}, A. 2006, \araa, 44, 141

\bibitem[{{Ruiz-Lara} {et~al.}(2015){Ruiz-Lara}, {P{\'e}rez}, {Gallart},
  {Alloin}, {Monelli}, {Koleva}, {Pompei}, {Beasley},
  {S{\'a}nchez-Bl{\'a}zquez}, {Florido}, {Aparicio}, {Fleurence}, {Hardy},
  {Hidalgo}, \& {Raimann}}]{ruizlara2015}
{Ruiz-Lara}, T., {P{\'e}rez}, I., {Gallart}, C., {et~al.} 2015, \aap, 583, A60

\bibitem[{{S{\'a}nchez-Bl{\'a}zquez} {et~al.}(2011){S{\'a}nchez-Bl{\'a}zquez},
  {Ocvirk}, {Gibson}, {P{\'e}rez}, \& {Peletier}}]{sanchezblazquez2011}
{S{\'a}nchez-Bl{\'a}zquez}, P., {Ocvirk}, P., {Gibson}, B.~K., {P{\'e}rez}, I.,
  \& {Peletier}, R.~F. 2011, \mnras, 415, 709

\bibitem[{{S{\'a}nchez-Bl{\'a}zquez} {et~al.}(2006){S{\'a}nchez-Bl{\'a}zquez},
  {Peletier}, {Jim{\'e}nez-Vicente}, {Cardiel}, {Cenarro},
  {Falc{\'o}n-Barroso}, {Gorgas}, {Selam}, \& {Vazdekis}}]{sanchez2006}
{S{\'a}nchez-Bl{\'a}zquez}, P., {Peletier}, R.~F., {Jim{\'e}nez-Vicente}, J.,
  {et~al.} 2006, \mnras, 371, 703

\bibitem[{{Sanders} {et~al.}(2007){Sanders}, {Salvato}, {Aussel}, {Ilbert},
  {Scoville}, {Surace}, {Frayer}, {Sheth}, {Helou}, {Brooke}, {Bhattacharya},
  {Yan}, {Kartaltepe}, {Barnes}, {Blain}, {Calzetti}, {Capak}, {Carilli},
  {Carollo}, {Comastri}, {Daddi}, {Ellis}, {Elvis}, {Fall}, {Franceschini},
  {Giavalisco}, {Hasinger}, {Impey}, {Koekemoer}, {Le F{\`e}vre}, {Lilly},
  {Liu}, {McCracken}, {Mobasher}, {Renzini}, {Rich}, {Schinnerer}, {Shopbell},
  {Taniguchi}, {Thompson}, {Urry}, \& {Williams}}]{sanders2007}
{Sanders}, D.~B., {Salvato}, M., {Aussel}, H., {et~al.} 2007, \apjs, 172, 86

\bibitem[{{Santini} {et~al.}(2014){Santini}, {Maiolino}, {Magnelli}, {Lutz},
  {Lamastra}, {Li Causi}, {Eales}, {Andreani}, {Berta}, {Buat}, {Cooray},
  {Cresci}, {Daddi}, {Farrah}, {Fontana}, {Franceschini}, {Genzel}, {Granato},
  {Grazian}, {Le Floc'h}, {Magdis}, {Magliocchetti}, {Mannucci}, {Menci},
  {Nordon}, {Oliver}, {Popesso}, {Pozzi}, {Riguccini}, {Rodighiero}, {Rosario},
  {Salvato}, {Scott}, {Silva}, {Tacconi}, {Viero}, {Wang}, {Wuyts}, \&
  {Xu}}]{santini2014}
{Santini}, P., {Maiolino}, R., {Magnelli}, B., {et~al.} 2014, \aap, 562, A30

\bibitem[{{Saracco} {et~al.}(2011){Saracco}, {Longhetti}, \&
  {Gargiulo}}]{saracco2011}
{Saracco}, P., {Longhetti}, M., \& {Gargiulo}, A. 2011, \mnras, 412, 2707

\bibitem[{{Scarlata} {et~al.}(2007){Scarlata}, {Carollo}, {Lilly}, {Sargent},
  {Feldmann}, {Kampczyk}, {Porciani}, {Koekemoer}, {Scoville}, {Kneib},
  {Leauthaud}, {Massey}, {Rhodes}, {Tasca}, {Capak}, {Maier}, {McCracken},
  {Mobasher}, {Renzini}, {Taniguchi}, {Thompson}, {Sheth}, {Ajiki}, {Aussel},
  {Murayama}, {Sanders}, {Sasaki}, {Shioya}, \& {Takahashi}}]{scarlata2007}
{Scarlata}, C., {Carollo}, C.~M., {Lilly}, S., {et~al.} 2007, \apjs, 172, 406

\bibitem[{{Schiavon}(2007)}]{schiavon2007}
{Schiavon}, R.~P. 2007, \apjs, 171, 146

\bibitem[{{Schlegel} {et~al.}(1998){Schlegel}, {Finkbeiner}, \&
  {Davis}}]{schlegel1998}
{Schlegel}, D.~J., {Finkbeiner}, D.~P., \& {Davis}, M. 1998, \apj, 500, 525

\bibitem[{{Scoville} {et~al.}(2007){Scoville}, {Aussel}, {Brusa}, {Capak},
  {Carollo}, {Elvis}, {Giavalisco}, {Guzzo}, {Hasinger}, {Impey}, {Kneib},
  {LeFevre}, {Lilly}, {Mobasher}, {Renzini}, {Rich}, {Sanders}, {Schinnerer},
  {Schminovich}, {Shopbell}, {Taniguchi}, \& {Tyson}}]{scoville2007}
{Scoville}, N., {Aussel}, H., {Brusa}, M., {et~al.} 2007, \apjs, 172, 1

\bibitem[{{Szomoru} {et~al.}(2011){Szomoru}, {Franx}, {Bouwens}, {van Dokkum},
  {Labb{\'e}}, {Illingworth}, \& {Trenti}}]{szomoru2011}
{Szomoru}, D., {Franx}, M., {Bouwens}, R.~J., {et~al.} 2011, \apjl, 735, L22

\bibitem[{{Tacchella} {et~al.}(2015{\natexlab{a}}){Tacchella}, {Carollo},
  {Renzini}, {Schreiber}, {Lang}, {Wuyts}, {Cresci}, {Dekel}, {Genzel},
  {Lilly}, {Mancini}, {Newman}, {Onodera}, {Shapley}, {Tacconi}, {Woo}, \&
  {Zamorani}}]{tacchella2015b}
{Tacchella}, S., {Carollo}, C.~M., {Renzini}, A., {et~al.} 2015{\natexlab{a}},
  Science, 348, 314

\bibitem[{{Tacchella} {et~al.}(2015{\natexlab{b}}){Tacchella}, {Lang},
  {Carollo}, {F{\"o}rster Schreiber}, {Renzini}, {Shapley}, {Wuyts}, {Cresci},
  {Genzel}, {Lilly}, {Mancini}, {Newman}, {Tacconi}, {Zamorani}, {Davies},
  {Kurk}, \& {Pozzetti}}]{tacchella2015a}
{Tacchella}, S., {Lang}, P., {Carollo}, C.~M., {et~al.} 2015{\natexlab{b}},
  \apj, 802, 101

\bibitem[{{Taylor} {et~al.}(2010){Taylor}, {Franx}, {Brinchmann}, {van der
  Wel}, \& {van Dokkum}}]{taylor2010}
{Taylor}, E.~N., {Franx}, M., {Brinchmann}, J., {van der Wel}, A., \& {van
  Dokkum}, P.~G. 2010, \apj, 722, 1

\bibitem[{{Thielemann} {et~al.}(1996){Thielemann}, {Nomoto}, \&
  {Hashimoto}}]{thielemann1996}
{Thielemann}, F.-K., {Nomoto}, K., \& {Hashimoto}, M.-A. 1996, \apj, 460, 408

\bibitem[{{Thomas} \& {Maraston}(2003)}]{thomas2003a}
{Thomas}, D., \& {Maraston}, C. 2003, \aap, 401, 429

\bibitem[{{Thomas} {et~al.}(2003){Thomas}, {Maraston}, \&
  {Bender}}]{thomas2003b}
{Thomas}, D., {Maraston}, C., \& {Bender}, R. 2003, \mnras, 343, 279

\bibitem[{{Thomas} {et~al.}(2005){Thomas}, {Maraston}, {Bender}, \& {Mendes de
  Oliveira}}]{thomas2005}
{Thomas}, D., {Maraston}, C., {Bender}, R., \& {Mendes de Oliveira}, C. 2005,
  \apj, 621, 673

\bibitem[{{Thomas} {et~al.}(2011){Thomas}, {Maraston}, \&
  {Johansson}}]{thomas2011}
{Thomas}, D., {Maraston}, C., \& {Johansson}, J. 2011, \mnras, 412, 2183

\bibitem[{{Thomas} {et~al.}(2004){Thomas}, {Maraston}, \& {Korn}}]{thomas2004}
{Thomas}, D., {Maraston}, C., \& {Korn}, A. 2004, \mnras, 351, L19

\bibitem[{{Thomas} {et~al.}(2010){Thomas}, {Maraston}, {Schawinski}, {Sarzi},
  \& {Silk}}]{thomas2010}
{Thomas}, D., {Maraston}, C., {Schawinski}, K., {Sarzi}, M., \& {Silk}, J.
  2010, \mnras, 404, 1775

\bibitem[{{Toft} {et~al.}(2007){Toft}, {van Dokkum}, {Franx}, {Labbe},
  {F{\"o}rster Schreiber}, {Wuyts}, {Webb}, {Rudnick}, {Zirm}, {Kriek}, {van
  der Werf}, {Blakeslee}, {Illingworth}, {Rix}, {Papovich}, \&
  {Moorwood}}]{toft2007}
{Toft}, S., {van Dokkum}, P., {Franx}, M., {et~al.} 2007, \apj, 671, 285

\bibitem[{{Trager} {et~al.}(2000){Trager}, {Faber}, {Worthey}, \&
  {Gonz{\'a}lez}}]{trager2000}
{Trager}, S.~C., {Faber}, S.~M., {Worthey}, G., \& {Gonz{\'a}lez}, J.~J. 2000,
  \aj, 120, 165

\bibitem[{{Trager} {et~al.}(1998){Trager}, {Worthey}, {Faber}, {Burstein}, \&
  {Gonzalez}}]{trager1998}
{Trager}, S.~C., {Worthey}, G., {Faber}, S.~M., {Burstein}, D., \& {Gonzalez},
  J.~J. 1998, \apjs, 116, 1

\bibitem[{{Trager} {et~al.}(2005){Trager}, {Worthey}, {Faber}, \&
  {Dressler}}]{trager2005}
{Trager}, S.~C., {Worthey}, G., {Faber}, S.~M., \& {Dressler}, A. 2005, \mnras,
  362, 2

\bibitem[{{Trujillo} {et~al.}(2007){Trujillo}, {Conselice}, {Bundy}, {Cooper},
  {Eisenhardt}, \& {Ellis}}]{trujillo2007}
{Trujillo}, I., {Conselice}, C.~J., {Bundy}, K., {et~al.} 2007, \mnras, 382,
  109

\bibitem[{{van de Sande} {et~al.}(2013){van de Sande}, {Kriek}, {Franx}, {van
  Dokkum}, {Bezanson}, {Bouwens}, {Quadri}, {Rix}, \&
  {Skelton}}]{vandesande2013}
{van de Sande}, J., {Kriek}, M., {Franx}, M., {et~al.} 2013, apj, 771, 85

\bibitem[{{van der Wel} {et~al.}(2009){van der Wel}, {Bell}, {van den Bosch},
  {Gallazzi}, \& {Rix}}]{vanderwel2009}
{van der Wel}, A., {Bell}, E.~F., {van den Bosch}, F.~C., {Gallazzi}, A., \&
  {Rix}, H.-W. 2009, \apj, 698, 1232

\bibitem[{{van der Wel} {et~al.}(2014){van der Wel}, {Franx}, {van Dokkum},
  {Skelton}, {Momcheva}, {Whitaker}, {Brammer}, {Bell}, {Rix}, {Wuyts},
  {Ferguson}, {Holden}, {Barro}, {Koekemoer}, {Chang}, {McGrath},
  {H{\"a}ussler}, {Dekel}, {Behroozi}, {Fumagalli}, {Leja}, {Lundgren},
  {Maseda}, {Nelson}, {Wake}, {Patel}, {Labb{\'e}}, {Faber}, {Grogin}, \&
  {Kocevski}}]{vanderwel2014}
{van der Wel}, A., {Franx}, M., {van Dokkum}, P.~G., {et~al.} 2014, \apj, 788,
  28

\bibitem[{{van Dokkum} {et~al.}(2008){van Dokkum}, {Franx}, {Kriek}, {Holden},
  {Illingworth}, {Magee}, {Bouwens}, {Marchesini}, {Quadri}, {Rudnick},
  {Taylor}, \& {Toft}}]{vandokkum2008}
{van Dokkum}, P.~G., {Franx}, M., {Kriek}, M., {et~al.} 2008, \apjl, 677, L5

\bibitem[{{Villumsen}(1983)}]{villumsen1983}
{Villumsen}, J.~V. 1983, \mnras, 204, 219

\bibitem[{{Whitaker} {et~al.}(2012){Whitaker}, {Kriek}, {van Dokkum},
  {Bezanson}, {Brammer}, {Franx}, \& {Labb{\'e}}}]{whitaker2011}
{Whitaker}, K.~E., {Kriek}, M., {van Dokkum}, P.~G., {et~al.} 2012, \apj, 745,
  179

\bibitem[{{Williams} {et~al.}(2009){Williams}, {Quadri}, {Franx}, {van Dokkum},
  \& {Labb{\'e}}}]{williams2009}
{Williams}, R.~J., {Quadri}, R.~F., {Franx}, M., {van Dokkum}, P., \&
  {Labb{\'e}}, I. 2009, \apj, 691, 1879

\bibitem[{{Williams} {et~al.}(2010){Williams}, {Quadri}, {Franx}, {van Dokkum},
  {Toft}, {Kriek}, \& {Labb{\'e}}}]{williams2010}
{Williams}, R.~J., {Quadri}, R.~F., {Franx}, M., {et~al.} 2010, \apj, 713, 738

\bibitem[{{Woosley} \& {Weaver}(1995)}]{woosley1995}
{Woosley}, S.~E., \& {Weaver}, T.~A. 1995, \apjs, 101, 181

\bibitem[{{Worthey} {et~al.}(1994){Worthey}, {Faber}, {Gonzalez}, \&
  {Burstein}}]{worthey1994}
{Worthey}, G., {Faber}, S.~M., {Gonzalez}, J.~J., \& {Burstein}, D. 1994,
  \apjs, 94, 687

\bibitem[{{Worthey} \& {Ottaviani}(1997)}]{worthey1997}
{Worthey}, G., \& {Ottaviani}, D.~L. 1997, \apjs, 111, 377

\bibitem[{{Wuyts} {et~al.}(2007){Wuyts}, {Labb{\'e}}, {Franx}, {Rudnick}, {van
  Dokkum}, {Fazio}, {F{\"o}rster Schreiber}, {Huang}, {Moorwood}, {Rix},
  {R{\"o}ttgering}, \& {van der Werf}}]{wuyts2007}
{Wuyts}, S., {Labb{\'e}}, I., {Franx}, M., {et~al.} 2007, \apj, 655, 51

\bibitem[{{Yano} {et~al.}(2016){Yano}, {Kriek}, {van der Wel}, \&
  {Whitaker}}]{yano2016}
{Yano}, M., {Kriek}, M., {van der Wel}, A., \& {Whitaker}, K.~E. 2016, \apjl,
  817, L21

\bibitem[{{Zanella} {et~al.}(2016){Zanella}, {Scarlata}, {Corsini}, {Bedregal},
  {Dalla Bont{\`a}}, {Atek}, {Bunker}, {.~Colbert}, {Dai}, {Henry}, {Malkan},
  {Martin}, {Rafelski}, {Rutkowski}, {Siana}, \& {Teplitz}}]{zanella2016}
{Zanella}, A., {Scarlata}, C., {Corsini}, E.~M., {et~al.} 2016, \apj, 824, 68

\end{thebibliography}

\end{document}